\documentclass[aps,amsmath,twocolumn,amssymb,floatfix,showpacs,superscriptaddress,nofootinbib,longbibliography]{revtex4-1}
\usepackage{mathtools}
\usepackage{braket}
\usepackage[dvipsnames]{xcolor}
\usepackage{float}
\usepackage{subfigure}
\usepackage{graphics}
\usepackage{bm}
\usepackage{MnSymbol}
\usepackage{braket}
\usepackage{float}
\usepackage{tikz}
\usepackage{blkarray}
\usepackage{pgffor}
\usepackage{float}
\usepackage{silence}
\usepackage{underscore}
\usepackage[english]{babel}
\usepackage[colorlinks=true,linktoc=page,linkcolor=ForestGreen,citecolor=magenta,urlcolor=purple]{hyperref}

\mathchardef\mhyphen="2D 

\newcommand{\etal}{{\it et al.}}
\newcommand{\ie }{{\it i.e.,\,\,}}

\newcommand\bea{\begin{eqnarray}}
\newcommand\eea{\end{eqnarray}}
\newcommand\beq{\begin{equation}}  
\newcommand\eeq{\end{equation}}

\usepackage[normalem]{ulem}
\definecolor{lime}{HTML}{A6CE39}
\usepackage{sidecap,tikz}
\DeclareRobustCommand{\orcidicon}{\hspace{-1.0mm}
	\begin{tikzpicture}
		\draw[lime, fill=lime] (0.0,0.0) 
		circle [radius=0.15] 
		node[white] {{\fontfamily{qag}\selectfont \tiny \,ID}};
		\draw[white, fill=white] (-0.0525,0.095) 
		circle [radius=0.007];
	\end{tikzpicture}
	\hspace{-3.0mm}
}
\foreach \x in {A, ..., Z}{\expandafter\xdef\csname orcid\x\endcsname{\noexpand\href{https://orcid.org/\csname orcidauthor\x\endcsname}
		{\noexpand\orcidicon}}
}

\AtBeginDocument{%
	\newwrite\bibnotes
	\def\bibnotesext{Notes.bib}
	\immediate\openout\bibnotes=\jobname\bibnotesext
	\immediate\write\bibnotes{@CONTROL{REVTEX41Control}}
	\immediate\write\bibnotes{@CONTROL{%
			apsrev41Control,author="08",editor="1",pages="1",title="1",year="1"}}
	\if@filesw
	\immediate\write\@auxout{\string\citation{apsrev41Control}}%
	\fi
}%
\begin{document}
\title{Spin polarization and diode effect in thermoelectric current through altermagnet-based superconductor heterostructures}

\author{Debika Debnath\orcidA{}}
\email{debika.uoh@gmail.com}
\affiliation{Theoretical Physics Division, Physical Research Laboratory, Navrangpura, Ahmedabad-380009, India}
\author{Arijit Saha\orcidC{}}
\email{	arijit@iopb.res.in}
\affiliation{Institute of Physics, Sachivalaya Marg, Bhubaneswar, Orissa 751005, India}
\affiliation{Homi Bhabha National Institute, Training School Complex, Anushakti Nagar, Mumbai 400094, India}
\author{Paramita Dutta\orcidB{}}
\email{paramita@prl.res.in}
\affiliation{Theoretical Physics Division, Physical Research Laboratory, Navrangpura, Ahmedabad-380009, India}
\begin{abstract}
The recent advent of a new class of magnetic material named as {\it altermagnet} (AM), characterized by a combination of momentum-dependent spin splitting with zero net magnetization, has opened up promising prospects for spintronic applications. We theoretically explore how the altermagnetic spin splitting affects the thermoelectric quasiparticle current in AM-based superconducting heterostructures. 
Our setup comprises of a bilayer system where a $d$-wave AM is proximity coupled to an ordinary $s$-wave superconductor (SC). We calculate the thermoelectric current carried by the quasiparticles applying a finite thermal bias across the junction. The behavior of the thermoelectric current with the system's base temperature and chemical potential is very similar to that in traditional SC heterostructures. Remarkably, the dissipative thermoelectric current found in the AM junction is spin split and thus generates finite spin polarization in the AM-based junction, which can approach $100\%$ spin polarization in the strong altermagnetic phase. We further investigate the thermoelectric current in AM-based Josephson junction (JJ) and illustrate how to achieve almost perfect diode effect in this AM-based JJ characterized by its efficiency $\sim 100\%$ with its sign decided by the strength of the AM, enhancing the potential for spin-caloritronics applications. 
\end{abstract}

\maketitle
\section{Introduction}
Thermoelectric phenomena couple voltage and temperature gradients through reciprocal effects, enabling energy harvesting~\cite{Callen1948,Harlingen1980,Lon2008,Goldsmid2009,Sothmann2014,Zhang2015,Goldsmid2017,Anatychuk2024}. These effects offer a deep insight into the fundamental physics resulting from the interplay of charge and heat transport. For over a century, significant research has been performed to investigate the thermoelectric properties of semiconductors and semimetals, which have been established as effective thermoelectric candidates~\cite{Domenicali1954,Domenicali1960,Brodsky1973,Barnard1974,Shakouri2011}. In recent decades, advances in material engineering have revealed novel heterostructures capable of exhibiting thermoelectric coefficients far exceeding those of normal metals or semiconductors. 
Although superconductors (SCs) had long been considered as speculative for higher thermoelectric performance, SCs with spin–orbit interaction or magnetic impurities or when integrated with other materials like ferromagnets possessing asymmetric spin-band structure have emerged as promising systems in this direction~\cite{Kalenkov2012,Bauer2012, Machon2013,Machon2014,Ozaeta2014,Giazotto2015,Hwang2016,Shelly2016,Kolenda2016,Dutta2017,Kolenda2017,Dutta2020}. The thermoelectric current in these systems has been shown to be higher when the symmetry of the energy distribution between electronlike and holelike quasiparticles is destroyed.

In recent times, a tremendous upsurge of research~\cite{Kolenda2016, Fornieri2017, Dutta2017, Kolenda2017, Pershoguba2019, Dutta2020, Muller2021, Dutta2023, Pal2024,  Zhang2024b,  Pan2025, Sano2025, Arrachea2025, Trocha2025, Mansikkamaki2025, Bera2025, Dutta2025, Arrachea2025} in the search for enhanced thermoelectricity has revealed excellent opportunities for quantum technologies. These range from spin caloritronics~\cite{Bauer2012, Linder2016, Keidel2020, Uchida2021, Yang2023} and realization of thermal analogs to electrical circuits~\cite{Lu2019, Majland2020, Tiwari2025}, to probing exotic quantum states~\cite{Wang2019, Hofmann2024, Smirnov2025} and potential schemes for thermally assisted quantum readout mechanisms via qubits~\cite{Gunyho2024}. 
This paves the way for future technological advances in quantum computation. Moreover, recent experimental advancements have provided more sophisticated paradigms to achieve fine tunability and phase-dependent thermoelectric current in Josephson junctions (JJs) enabling potential applications including thermoelectric quantum sensors via phase-controlled thermoelectric current~\cite{Guttman1997, Pershoguba2019, Marchegiani2020, Marchegiani2020b, Mukhopadhyay2022, Dutta2023, Chatterjee2024, Mansikkamaki2025}. 
However, full empirical validation of theoretical predictions remains challenging, primarily due to the difficulty in generating and controlling large spin splitting through high magnetic moments—an essential requirement for achieving strong spin-dependent thermoelectricity. Therefore, finding an efficient alternative route for enhanced thermoelectricity is one of the engrossing studies of modern condensed matter research.

The advent of a new magnetic phase, known as altermagmetism, provides a fertile ground for studying spin-dependent electronic transport properties due to its unconventional energy band structure with finite non-relativistic spin splitting in the absence of a net magnetization~\cite{Smejkal2022b, Mazin2022, Feng2022, Zhu2024, Song2025}. Altermagnets (AMs) are distinct from both ferromagnets and anti-ferromagnets due to their intrinsic crystal geometry with collinear and non-collinear structure~\cite{Bai2024, Cheong2024}. These materials have emerged as promising platforms to explore the time-reversal symmetry-broken phases without any stray magnetization~\cite{Fedchenko2024}, while their intrinsic spin configuration generates spin-dependent current. These have been shown to carry the potential to execute thermoelectric phenomena without any additional magnetic field~\cite{Sukhachov2024, Zhou2024b, Ashani2025, Yi2025}, thereby opening a new frontier for the spin-split thermoelectricity. Recent observation of anomalous Nernst effect in AM materials $\text{Mn}_{5}\text{Si}_{3}$~\cite{Badura2025} and CrSb single crystals~\cite{Li2025} shows the potential of AM in the same direction. 
On top of that, recent findings of highly controllable AM parameters~\cite{Gomonay2024, Smejkal2024, Duan2025, Gu2025} offer the opportunity to develop a tunable spin-dependent thermoelectric current.

Very recently, significant research has demonstrated that coupling AM to SC can produce finite momentum Cooper pairs without any magnetization~\cite{Ouassou2023, Zhang2024}. In two independent works, Papaj~\cite{Papaj2023} and Sun \etal~\cite{Sun2023}, investigated the spin-split supercurrent in AM-SC junctions. In addition to the electrical current, thermoelectric effect has also been studied by Sukhachov \etal~in the SC-AM heterojunction using the inverse proximity effect~\cite{Sukhachov2024}. This result manifests a nonmonotonic response of thermoelectric current in the AM-SC-AM junction, albeit in a spin-independent framework. However, the key characteristics of AM lie in their momentum-dependent spin splitting. This naturally leads to the immediate fundamental question: What is the impact of AM spin splitting on thermoelectric current in SC-AM heterostructures?

Moreover, the recent advent of unidirectional supercurrent in JJs gives rise to the phenomena of Josephson diode effect (JDE). The latter has become a focal point of dissipationless quantum transport research~\cite{Ando2020, Misaki2021,  Baumgartner2022, Davydova2022, Nagaosa2022, Zhang2022, Mao2024, Debnath2024, Debnath2025, Roy2025}. With the onset of AM materials, a few recent works have reported the JDE in AM-based JJs~\cite{Banerjee2024, Cheng2024, Sharma2025} in addition to the perfect diode effect in bare AM material~\cite{Chakraborty2025}. The control over the spin splitting plays the key role in tuning the nonreciprocity of the Cooper pair current. Additionally, the symmetry argument for JDE in AM-JJ is established to be different than regular JJ~\cite{Cheng2024, Banerjee2024, Sharma2025}. In parallel, the crucial role of thermal gradient-induced unidirectional currents in heat-based signal processing~\cite{Wehmeyer2017, Li2022} has fueled growing interest in studying nonreciprocal transport under finite thermal bias~\cite{Fornieri2014,  Martinez2015, Chatterjee2024, Goury2019, Antola2024, Balduque2025, Mercebach2025}. However, the nonreciprocity in thermoelectric current in AM based JJ still remains an open question.

Motivated by this, in this article, our primary goal is to find out the impact of AM spin splitting on the thermoelectric current in a minimal model of 
$d$-wave AM-regular $s$-wave SC junction and subsequently in phase-biased AM-based JJs to investigate the spin-polarized thermoelectric current. We systematically explore the dependence of the quasiparticle-mediated thermoelectric current on AM parameters, junction temperature, and chemical potential. Very recently, Chen \etal ~\cite{Chen2025} have reported the behavior of the thermal current in a thermally biased AM-based JJ. In contrast, we focus on the spin-dependent thermoelectric current mediated by the quasiparticles. Finally, we demonstrate the nonreciprocity behavior of the thermoelectric current in a AM-based JJ incorporating Rashba spin-orbit interaction (RSOI) and proximity-induced superconductivity in AMs.

We organize the remainder of the article as follows. In Sec.~\ref{bilayer}, we introduce the basic model of AM-SC bilayer and investigate the spin-dependent thermoelectric current for a range of AM parameters. Then, we explore an AM-based JJ in Sec.~\ref{SC_AM_SC:JJ} and examine the behavior of the thermoelecric current with respect to the SC phase difference in addition to the AM strengths, junction temperature, and chemical potential. In Sec.~\ref{SC_AM_SC:JD}, we investigate the emergence of nonreciprocal thermoelectric current in the modified AM-based JJ, satisfying the symmetry requirements and thus establish the \textit{diode effect in thermoelectric current in spin-split AM-based JJ}. 
Finally, we summarize and conclude our paper in Sec.~\ref{Summary}.

\section{Altermagnet-superconductor bilayer junction (AM-SC)}\label{bilayer}

\begin{figure}[h!]
\centering
\includegraphics[scale=0.45]{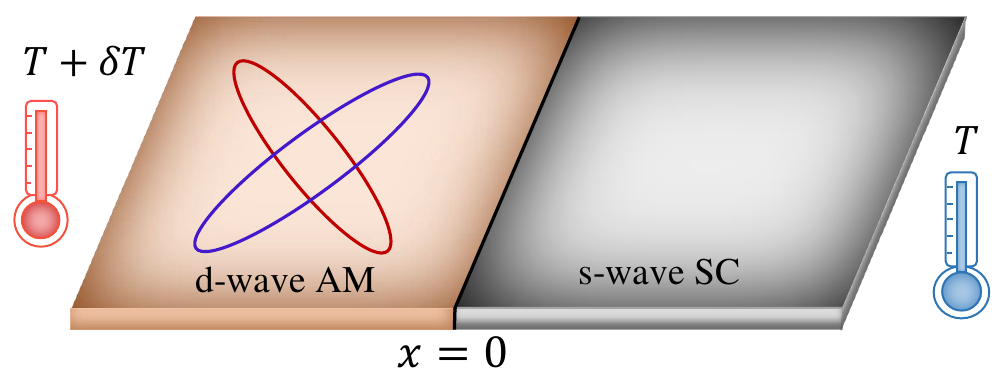}
\caption{Schematic diagram of thermally biased AM-SC hetero junction. Here, a $d$-wave AM is proximity coupled to a common bulk $s$-wave SC at $x=0$. A temperature gradient $\delta T$ is applied across the junction.}
\label{fig:amsc}
\end{figure}
To begin with, we study a AM-SC bilayer to explore the effect of AM spin splitting on the thermoelectric current through this geometry.
\subsection{Model and method}
We consider a two-dimensional ($2$D) minimal model as shown in Fig.~\ref{fig:amsc} where a $d$-wave AM is coupled to an ordinary $s$-wave superconductor 
at $x=0$ interface forming AM-SC heterojunction. We consider a thermal gradient applied across the junction setting the AM at a higher temperature $T+\delta T$ compared to the SC maintained at $T$. To describe this junction, we employ the Bogoliubov-deGennes (BdG) Hamiltonian in the Nambu basis $\psi_{\boldsymbol{k}}=(\psi_{k,\uparrow},\psi_{k,\downarrow}, \psi^{\dagger}_{-k,\uparrow},\psi^{\dagger}_{-k,\downarrow})^{T}$ as
\begin{equation}
\mathcal{H}=\frac{1}{2}\sum_{\boldsymbol{k}}\psi_{\boldsymbol{k}}^{\dagger} \mathcal{H}_{\boldsymbol{k}} \psi_{\boldsymbol{k}}\ ,
\end{equation}
where
\begin{equation}
\mathcal{H}_{\boldsymbol{k}}=\epsilon_{\boldsymbol{k}} \tau_{z}  \sigma_{0}+ t_{\boldsymbol{k},\sigma} \tau_{z} \sigma_{z}  - \Delta \tau_{y} \sigma_{y}\ ,
    \label{ham:bilayer}
\end{equation}
with $\boldsymbol{k}=(k_{x}, k_{y})$ and $\tau_{i}(\sigma_{i})$ (for $i=x,y,z$) denotes the Pauli matrices in particle-hole (spin) space. 
The kinetic energy of the quasiparticles is given by
\begin{eqnarray}
   \epsilon_{\boldsymbol{k}}&=&[t_{0}(k_{x}^{2}+k_{y}^{2})-\mu_{\rm AM}] \Theta(-x) \nonumber \\ && ~~~~ 
   +[t_{\rm SC}(k_{x}^{2}+k_{y}^{2})-\mu_{\rm SC}] \Theta(x)\ ,
   \label{ham:amsc}
\end{eqnarray}
where, $t_{0}$ and $t_{\rm SC}$ correspond to the hopping amplitudes of the AM and SC, respectively. The chemical potentials in the AM and SC are denoted by $\mu_{\rm AM}$ and $\mu_{\rm SC}$, respectively. We consider the $d$-wave AM described by~\cite{Papaj2023, Sun2023} 
\begin{eqnarray}
    t_{\boldsymbol{k},\sigma}& =& [t_{1} (k_{y}^{2}-k_{x}^{2}) +2t_{2} k_{x}k_{y}]\Theta(-x)\ . \label{ham:am}
\end{eqnarray}
The parameters $t_{1}$ and $t_{2}$ characterize the AM with different AM-SC interface orientations: (i) $t_{1}\neq 0$ and $t_{2}=0$ correspond to the $d_{x^2-y^2}$ symmetry of the AM favoring the normal incidence, (ii) $t_{1}=0$ and $t_{2}\neq 0$ describe $45^{\circ}$ rotation with $d_{xy}$ AM symmetry, and (iii) both finite $t_{1}, t_{2} \neq 0$ represent an arbitrary rotation of the AM-SC interface. The role of $t_1$ and $t_2$ (including their simultaneous presence) in the spin splitting can be understood in terms of the total density of states (DOS) of the AM as presented in Appendix~\ref{dos}. 

The temperature dependence of the gap parameter of the common $s$-wave SC reads $\Delta=\Delta_{0}\text{tanh}(1.74\sqrt{T_{c}/T-1})\Theta(x)$ with $T_c$ being the critical temperature of the SC. To protect the junction from the formation of the intrinsic barrier due to Fermi-wave vector asymmetry, we fix $\mu_{\rm AM}=\mu_{\rm SC}=\mu$. We also set $t_{0}=t_{\rm SC}$ throughout the article for the sake of simplicity. However, this choice does not affect our results qualitatively. All temperatures are scaled by the superconductor transition temperature $T_{c}$ and the chemical potentials are expressed in units of $\Delta_{0}$.

We calculate the thermoelectric current, i.e., the charge current in our thermally biased AM-SC hetrostructure. Within the linear response regime, the spin-resolved thermoelectric current by an electron with spin $\sigma$ is given by~\cite{Dutta2017} 
\begin{equation}
    \mathcal{L}_{\sigma}=\frac{2e}{h T} \int_{0}^{\infty} \left(E-\mu \right) \left(1-|r_{\text{N}_{e}}^{\sigma}|^{2}+|r_{\text{A}_{h}}^{\sigma}|^{2}\right) \left(-\frac{df(E,T)}{dE}\right) dE \ ,
    \label{L_bilayer}
\end{equation}
where $|r_{\text{N}_{e}}^{\sigma}|^{2}$ and $|r_{\text{A}_{h}}^{\sigma}|^{2}$ represent the normal and Andreev reflection (AR) probabilities for an incident spin $\sigma$ electron at the interface of the AM-SC heterojunction, respectively. It is important to note that due to particle-hole symmetry, the current carried by the incident electron or hole are the same for the bilayer structure. Here $f(E,T)$ represents the Fermi distribution function $f(E,T)=(\text{Exp}[(E-\mu)/k_B T]+1)^{-1}$. The term $\mathcal{L}_{\sigma}$ denotes the off-diagonal element of the Onsager's matrix mentioned in Appendix~\ref{onsagar} that describes both conjugate and nonconjugate processes connecting the responses of the voltage-biased and temperature-biased systems. The current in Eq.~\eqref{L_bilayer} is expressed in terms of the scattering amplitudes that can be modified by the Fermi-energy distributions and the energy measured with respect to the chemical potential of the system. To calculate the reflection and AR coefficients in our AM-SC bilayer, we employ the scattering matrix formalism which is outlined in  Appendix~\ref{scatmat}. We use the boundary conditions described in Eqs.~\eqref{bc1} and \eqref{bc2} and numerically solve them simultaneously to obtain the scattering coefficients, which enter into the current expression [Eq.~\eqref{L_bilayer}] to obtain the thermoelectric current. We use the amplitudes to calculate the spin-dependent thermoelectric current $\mathcal{L}_{\sigma}$, i.e., $\mathcal{L}_{\uparrow}$ and $\mathcal{L}_{\downarrow}$.

\subsection{Results and discussions}
Here, we present our results for the thermoelectric current in AM-SC junction and show how the thermoelectric current is regulated by the AM strengths, junction temperature, and the chemical potential.

\subsubsection{Effect of AM parameters}
\begin{figure}
\centering
\includegraphics[scale=0.485]{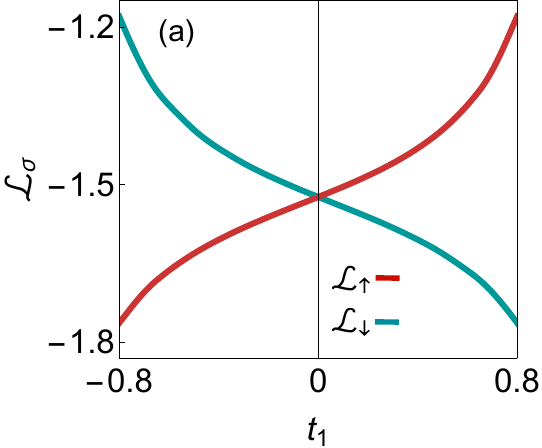}\hspace{-4mm}
\includegraphics[scale=0.485]{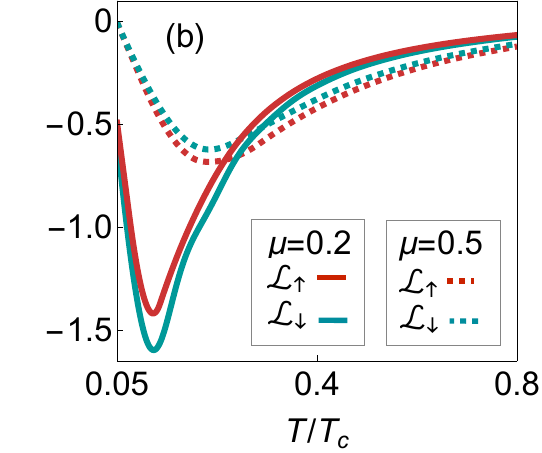}
\caption{The behavior of the spin-split thermoelectric current $\mathcal{L}_{\sigma}$ (in units of $ek_{B}/h$) as a function of (a) $t_{1}$ choosing $t_{2}=0$ and (b) junction temperature $T/T_{c}$. The other model parameters are considered as (a) $\mu=200\Delta_0$, $T=0.1T_{c}$, $\Delta_0=0.001$ and (b) $t_{1}=0.4$, $t_{2}=0$, $\mu=200\Delta_0$ and $500\Delta_0$ where $\Delta_0=0.001$.}
\label{L_vs_t:bilayer}
\end{figure}
In case of the AM-SC hybrid structure, an electron incident from the AM to the interface ($x=0$) may reflect back due to the ordinary reflection at the junction. Additionally, due to the AR process, an electron with spin $\sigma$ can also reflect back from the interface as a hole with spin $\Bar{\sigma}$, and a complementary Cooper pair transmits through the SC. The scattering properties of the junction are characterized by these two types of reflections and control the flow of Cooper pair and quasiparticles in the system below and above the SC gap. 

We consider up- and down-spin quasiparticles (electronlike and holelike) originating from the AM side as the incident particle within our AM-SC junction to calculate the thermoelectric currents $\mathcal{L}_{\uparrow}$ and $\mathcal{L}_{\downarrow}$, respectively. There are two AM parameters available: $t_{1}$ and $t_{2}$. Hence, we show the currents as a function of $t_{1}$ in Fig.~\ref{L_vs_t:bilayer}(a) by keeping $t_{2}=0$. Note that, throughout our analysis, we mention the amplitude of the current only (ignoring the sign) since the current being carried by electrons only, is always negative throughout our study. For $t_1>0$, we observe that $\mathcal{L}_{\downarrow}$ increases with the AM strength, whereas, the current $\mathcal{L}_{\uparrow}$ due to the up-spin incident particles reduces with increasing $t_{1}$. Moreover, with the increase in the AM strength $t_{1}$, the difference between $\mathcal{L}_{\uparrow}$ and $\mathcal{L}_{\downarrow}$ enhances since the spin splitting effect becomes more prominent as the system enters into the stronger AM phase. This spin splitting effect still carries it's signature in the behavior of DOS of the AM presented in Appendix~\ref{dos}. Here, the key point to mention is that the AM strength can be tuned to negative values, which effectively rotates the Fermi surface of the AM. As a result, we obtain the opposite behavior of $\mathcal{L}_{\sigma}$ at $t_{1}<0$ regime compared to our results for $t_{1}>0$. This is also realizable from the AM Hamiltonian of Eq.~(\ref{ham:am}). Although spin-split thermoelectric currents have been discussed only for $t_2=0$, we confirm that the presence of finite $t_{2}$ does not affect the behavior of $\mathcal{L}_{\sigma}$ qualitatively. Therefore, only the term $t_{1} (k_{y}^{2}-k_{x}^{2})$ of the AM Hamiltonian is responsible for the momentum-dependent spin splitting in the AM-SC bilayer. Note that, the role of the \textit{non-spin splitting parameter $t_{2}$} is already reported in the thermal current study of AM-JJ in terms of the orientation angle of AM-SC interface~\cite{Chen2025}. It is important to mention that we consider zero potential barrier strength at the interface(s) of our heterostructure(s) throughout the manuscript. As the barrier potential strength is enhanced, the current is reduced monotonically, as explained in Appendix~\ref{appen:amsc_Z}.

\subsubsection{Effect of temperature and chemical potential}\label{appen:L_bilayer}
 We find that the response of $\mathcal{L}_{\sigma}$ in an AM-SC junction is similar to that in a normal-SC junction concerning the temperature and chemical potential, except for the spin splitting behavior. The quasiparticles corresponding to the strong coherence peaks at the edges or just above of the SC gap in the energy band structure primarily contribute to the thermoelectricity. Thermal excitations of these quasiaparticles cause the enhancement of the thermoelectric current at low temperatures. After a certain temperature, the SC gap $\Delta$ drops down. Although there are more thermally excited quasiparticles at higher temperatures, the thermoelectric response of the bilayer junction becomes weaker at higher temperatures because of the symmetry in the energy band structure with the sharp coherence peaks getting smeared out. For higher temperatures, the difference between the two spin-split thermoelectric currents also decreases, as shown in Fig.~\ref{L_vs_t:bilayer}(b). We also find that $\mathcal{L}_{\sigma}$ reduces with the increase in the chemical potential $\mu$ of the heterostructure. Additionally, the shift observed in Fig.~\ref{L_vs_t:bilayer}(b) for the higher $\mu$ is justified mathematically due to the presence of the term $(E-\mu)(1/T) (\partial f/\partial E)$ in Eq.~(\ref{L_bilayer}). For the parameter regime $t_{1}=0.4$, the down-spin quasiparticle current $\mathcal{L}_{\downarrow}$ dominates over the up-spin current $\mathcal{L}_{\uparrow}$, as observed in Fig.~\ref{L_vs_t:bilayer}(a) at $\mu=200\Delta_{0}$. The change in $\mu$ can change the spin dominance on the spin split thermoelectric current as observed in Fig.~\ref{L_vs_t:bilayer}(b). Specifically, at $\mu=0.2$, the down-spin current dominates, whereas the up-spin current is enhanced when $\mu$ is higher. Therefore, it is established that an AM can serve as a good candidate for the generation of the spin-dependent thermoelectric current in an SC hybrid junction. Note that, we confirm the primary contributions to the current in our AM-SC setup are arising from quasiparticles by breaking the integration limit of Eq.~(\ref{L_bilayer}) in two parts, consisting of only the SC gap and all energy levels above the SC gap.

\section{Altermagnet-Josephson Junction (AM-JJ)}\label{SC_AM_SC:JJ}

In this section, we focus on the spin-dependent thermoelectric current in AM-JJ, where in addition to the AM parameters $t_{1}$ and $t_{2}$, SC phases 
also play a key role.

\subsection{Model and method}
\begin{figure}
\centering
\includegraphics[scale=0.4]{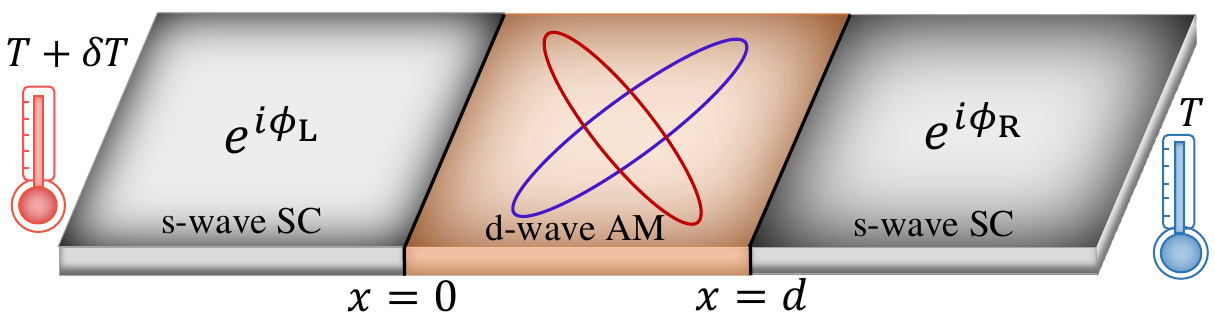}
\caption{A schematic diagram of AM-JJ setup where two $s$-wave SCs with phases $\phi_{\rm{L}}$ and $\phi_{\rm{R}}$ are coupled to a $d$-wave AM at $x=0$ and $x=d$ respectively. These two SCs are maintained at two different temperatures $T+\delta T$ and $T$, respectively, in order to create a temperature gradient $\delta T$ across the JJ setup.}
\label{fig:SC_AM_SC}
\end{figure}

We refer to Fig.~\ref{fig:SC_AM_SC} for the schematic diagram of the AM-JJ where two ordinary $s$-wave superconductors are coupled to a $d$-wave AM. Two SCs 
are maintained at two different temperatures to introduce a temperature gradient in the system along with a SC phase bias. The model Hamiltonian of the JJ 
can be expressed using the Heaviside theta function as,
\begin{equation}
     \mathcal{H}^{\text{JJ}}=\mathcal{H}_{\rm L}\Theta(-x)+\mathcal{H}_{\rm AM}\Theta(x) \Theta(d-x)+\mathcal{H}_{\rm R}\Theta(x-d)\ ,
    \label{Ham:JJ}
\end{equation}
where $\mathcal{H}_{\rm L}$ and $\mathcal{H}_{\rm R}$ represent the left and right SCs located at $x\le 0$ and $x\ge d$ regions, respectively. The three parts of the Hamiltonian corresponding to the individual AM and SCs are similar to Eq.~(\ref{ham:bilayer}) and finally, it looks like
\begin{equation}
    \mathcal{H}^{\text{JJ}}=\epsilon^{\text{JJ}}_{\boldsymbol{k}} \hat{\tau}_{z} \hat{\sigma}_{0} + t^{\text{JJ}}_{\boldsymbol{k},\sigma}\hat{\tau}_{z}\hat{\sigma}_{z}  -\Delta^{\text{JJ}}\hat{\tau}_{y}  \hat{\sigma}_{y}\ ,  
    \label{ham:JJ}
\end{equation}
where,
\begin{eqnarray}
    \epsilon^{\text{JJ}}_{\boldsymbol{k}}&=&[t_{0}(k_{x}^{2}+k_{y}^{2})-\mu_{\text{AM}}] \Theta(x)\Theta(d-x) \nonumber \\
     && +[t_{\text{SC}}(k_{x}^{2}+k_{y}^{2})-\mu_{\text{SC}}] \Theta(-x) \Theta(x-d)\ ,\\
     t^{\text{JJ}}_{\boldsymbol{k},\sigma}& = &[t_{1} (k_{y}^{2}-k_{x}^{2})+2t_{2} k_{x}k_{y}]\Theta(x) \Theta(d-x)\ ,\label{Ham:AM_JJ}\\
     \Delta^{\text{JJ}}&=&\Delta_{\text{L}} \Theta(-x)+\Delta_{\text{R}} \Theta(x-d)\ .
     \label{eq:AM_JJ}
\end{eqnarray}
Here, the two SCs are distinguished by their superconducting gap parameters, i.e., $\Delta_{\text{L(R)}}=\Delta_{0} e^{i \phi_{\text{L(R)}}} \text{tanh}(1.74\sqrt{T_{c}/T_{\text{L(R)}}-1})$, where we consider the SC phase for the left lead as $\phi_{\text{L}}=\phi/2$ and for the right lead as $\phi_{\text{R}}=-\phi/2$, resulting in a net SC phase difference of $\phi_{\text{L}}-\phi_{\text{R}}=\phi$ in the JJ. To avoid any effective barrier (density mismatch) to be established at the junction due to the differences in the AM and SC chemical potentials, we maintain $\mu_{\text{AM}}=\mu_{\text{SC}}=\mu \gg\Delta_{0}$ within the Andreev approximation. All other model parameters carry the same description as mentioned for the AM-SC setup in Sec.~\ref{bilayer}.

In the absence of any temperature gradient, \ie~at $\delta T=0$, the thermal equilibrium results in the exactly equal thermoelectric currents in the two individual SC leads, which effectively results in the zero net thermoelectric current in the JJ. As soon as a finite temperature gradient is applied, the quasiparticle current flows from the higher thermal gradient region towards the lower temperature domain. The total charge current for a particular thermal bias $\delta T$ can be calculated as the difference between the currents in the two opposite leads. This finally produces a thermally driven quasiparticle current given by~\cite{Pershoguba2019, Mukhopadhyay2022, Dutta2023},
\begin{equation}
    \mathcal{L}_{\sigma}=\frac{2e}{hT} \int_{\Delta_{\text{max}}}^{\infty}\frac{(E-\mu)}{\sqrt{E^2-\Delta_{\text{L(R)}}^2}} (i^{e}_{\sigma}-i^{h}_{\sigma}) \frac{df(E,T)}{dE} dE\ ,
     \label{L12}
\end{equation}
where $i^{e}_{\sigma}$ and $i^{h}_{\sigma}$, respectively denote the currents due to the electronlike and holelike quasiparticles with spin $\sigma$, incident from the SC towards the AM region, and $\Delta_{\text{max}}=\text{max}(\Delta_{\text{L}},\Delta_{\text{R}})$. The individual quasiparticle currents can be written as $i^{e(h)}_{\sigma}=1-|r_{\text{N}_{e(h)}}^{\sigma}|^{2}+|r_{\text{A}_{h(e)}}^{\sigma}|^{2}+\frac{2\Delta}{E}\text{Re}[r_{\text{A}_{h(e)}}^{\sigma}]$ and further can be expressed as $i^{e}_{\text{L}}=|t_{ee}^{\sigma}|^{2}-|t_{he}^{\sigma}|^{2}$ and $i^{h}_{\text{L}}=|t_{hh}^{\sigma}|^{2}-|t_{eh}^{\sigma}|^{2}$~\footnote{This is true when $2uv=\Delta/\epsilon_{k}$ by neglecting the term $\propto \epsilon_{k}/E_{F}$. See Appendix~\ref{scatmat:JJ} for the description of these terms and Ref.~[\onlinecite{Pershoguba2019}] for more details.}. The scattering coefficient $|t_{ee(hh)}^{\sigma}|^{2}$ denotes the normal transmission probability for an incident electronlike (holelike) quasiparticle and $|t_{he(eh)}^{\sigma}|^{2}$ indicates to the transmission probability corresponding to the opposite charge particles. To calculate the spin-split thermoelectric particle current $\mathcal{L}_{\sigma}$ for the AM-JJ, 
we employ the scattering matrix formalism and calculate the transmission coefficients using the appropriate boundary conditions. We set the temperatures 
in the two leads as $T_{\text{L}}=T + \delta T$ and $T_{\text{R}}=T$  to find the total thermoelectric current in our JJ. Note that, for the quasiparticles 
with energy $E\geq\Delta_{0}$, we find $t_{he}^{\sigma}=t_{eh}^{\sigma}\approx 0$. We refer to Appendix~\ref{scatmat:JJ} for the details. 
\begin{figure}
\centering
\hspace{-2.5mm}
\includegraphics[width=0.498\linewidth]{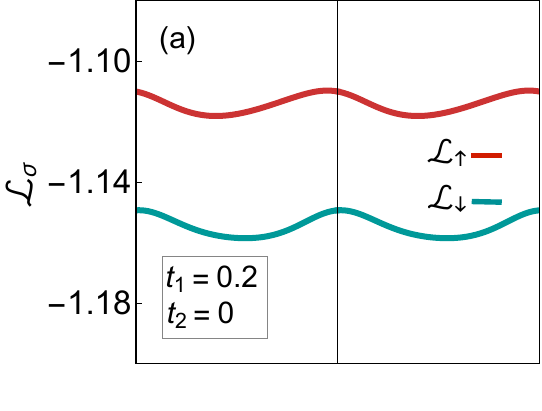} 
\includegraphics[width=0.485\linewidth]{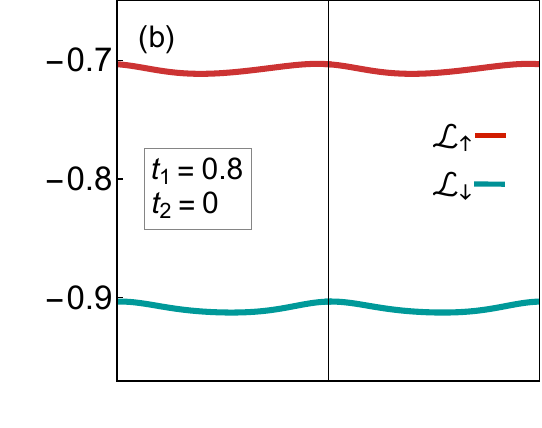}\\ \vspace{-2mm} 
\hspace{-0.5mm}
\includegraphics[width=0.4845\linewidth]{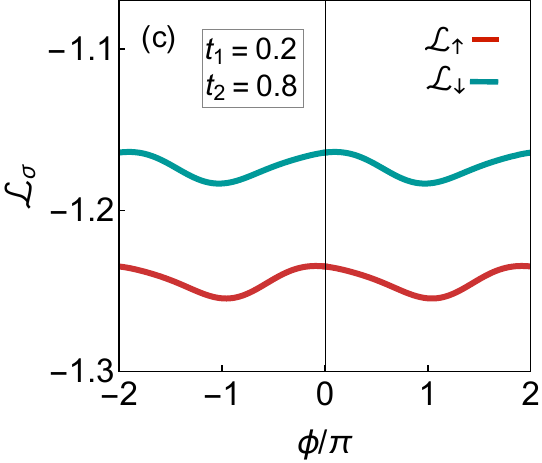} \hspace{-1.5mm}
\includegraphics[width=0.4845\linewidth]{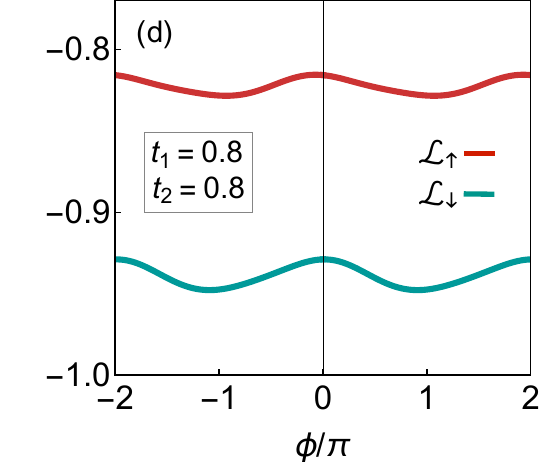}\\
\vspace{0.5cm}\hspace{1cm}
\caption{The spin-dependent thermoelectric quasiparticle currents $\mathcal{L}_{\sigma}$ (in units of $ek_{B}/h$) with respect to the superconducting phase difference $\phi$, for different pairs of AM strengths $t_{1}$ and $t_{2}$ in the JJ. The up-spin current is presented by $\mathcal{L}_{\uparrow}$ and the down-spin current is presented by $\mathcal{L}_{\downarrow}$. Here, we consider the other parameters of the model Hamiltonian as $\mu=200\Delta_{0}, d=0.1\xi, T=0.4 T_{c}, \Delta_{0}=0.001$.}
\label{L_vs_phi}
\end{figure}
\subsection{Results and discussions}
In this section, we discuss our results for the thermoelectric current in the AM-JJ setup and discuss how the thermoelectric current is regulated by the superconducting phase difference in addition to the AM strength, junction temperature, and chemical potential.

Before we move to the discussions on the current regulation by various parameters, a few remarks are in order. There are two distinct biases that are driving our AM-JJ. One is the SC phase difference between the two SC leads which causes the supercurrent flow through the junction mediated by the Cooper pairs. Another one is the finite thermal gradient that induces quasiparticle flow due to thermally excited carriers. For a short JJ where the junction length is smaller than or comparable to the superconducting phase coherence length, the quasiparticle current can be categorized into two components: (a) a nondissipative current arising from the coherent tunneling of the Cooper pairs, i.e., the conventional Josephson current, and (b) a dissipative current carried by the quasiparticles above the SC gap~\cite{Pershoguba2019}. Hence, a separation of the contributions by the Cooper pair and the quasiparticles is necessary to understand the behavior of the thermoelectric current. Since the origin of these two types of currents are different, and can be classified by the SC gap $\Delta_{\text{L(R)}}$, which is a function of the SC phase $\phi_{\text{L(R)}}$, the dissipative and nondissipative currents can be distinguished by reversing the SC phase difference $\phi$ in the JJ as follows. The total thermoelectric current $(\mathcal{L}=\mathcal{L}_{\uparrow}+\mathcal{L}_{\downarrow})$ can be expressed as $\mathcal{L}(\phi)=\mathcal{L}^{e}(\phi)+\mathcal{L}^{o}(\phi)$, where $\mathcal{L}^{e(o)}(\phi)$ represents the even (odd) thermoelectric current in the JJ with respect to $\phi$. They can be separated as $\mathcal{L}^{e}(\phi)=\mathcal{L}(\phi)+\mathcal{L}(-\phi)$ and $\mathcal{L}^{o}(\phi)=\mathcal{L}(\phi)-\mathcal{L}(-\phi)$. Here, the even part of the total current $\mathcal{L}^{e}$ represents the dissipative thermoelectric current by the quasiparticles, whereas the odd part $\mathcal{L}^{o}$ signifies the nondissipative usual Josephson current. Using the Landauer transport formalism~\cite{Emberly2020} (with the superconducting coherence length much greater than the Fermi wave vectors of the leads), we calculate the dissipative and nondissipative currents in our AM-JJ setup~\cite{Pershoguba2019} and confirm that the current carried in our AM-JJ is almost entirely dissipative; the even-in $\phi$ component of the total current dominates over the odd part being negligibly small. A detailed discussion regarding this can be found in Appendix~\ref{appen:evenodd}. Henceforth, all our discussions and conclusions made in this article are based on the dissipative thermoelectric currents since the nondissipative components of the currents are negligibly small. Note that, the length of the intermediate AM does not play any role in the spin splitting effect as it is the intrinsic property, whereas the chemical potential acts as a crucial parameter in the thermoelectric current study. We refer to Appendix~\ref{appen:L_vs_d_mu} for further details.

\subsubsection{Superconducting phase tunability and effect of AM parameters}
To reveal the tunability of the current by the SC phase difference, we depict the spin-resolved thermoelectric currents $\mathcal{L}_{\sigma}$ as a function of $\phi$  for various AM strengths in Fig.~\ref{L_vs_phi}. The behavior of the current profile is oscillating with respect to the SC phase difference with the amplitudes and the spin splitting largely modified by the AM parameters. In Figs.~\ref{L_vs_phi}(a) and ~\ref{L_vs_phi}(b) we present the thermoelectric current for the normal incidence ($t_2=0$) of the quasiparticles at the AM-SC interface of the JJ. We find that as the AM strength $t_{1}$ increases, the thermoelectric current reduces for both up- and down-spin particles, whereas the difference between the up and down spin currents becomes much more pronounced, as clear by comparing Figs.~\ref{L_vs_phi}(a) and ~\ref{L_vs_phi}(b). On the other hand, when we compare Figs.~\ref{L_vs_phi}(a) and ~\ref{L_vs_phi}(c), we find that for the same $t_{1}$ parameter as we switch on $t_{2}$ considering a rotated AM-SC interface of the JJ, the current is subsequently enhanced. It is important to notice that for $t_{2}>t_{1}$, the up-spin current is larger than the down-spin current. The reason behind this opposite spin dominance behavior lies in the variation of the momentum $k_{e(h)\uparrow}$ and $k_{e(h)\downarrow}$ 
(see Appendix~\ref{scatmat} for details) for $t_{1}>t_{2}$, which effectively alters the Fermi surface of the AM-SC interface. 
From Figs.~\ref{L_vs_phi}(b) and ~\ref{L_vs_phi}(d), we find a similar behavior as explained by comparing Fig.~\ref{L_vs_phi}(a) and (b).
Notably, although $t_{2}$ does help to generate the higher thermoelectric spin current, the difference between the two spin components of the current is enhanced in the presence of $t_{1}$. Thus, it is possible to tune the effective spin-dependent thermoelectric currents by modulating the AM parameters. Hence, it is more promising to see the variation of the spin-dependent current for a range of the AM rotational angle. We define the AM parameters as a function of AM rotational angle and discuss the corresponding thermoelectric current behavior in Appendix~\ref{appen:L_vs_theta}.

\begin{figure}
\centering
\includegraphics[width=0.495\linewidth]{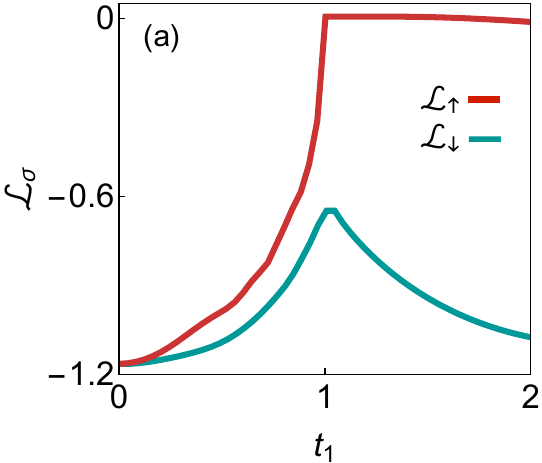}
\hspace{-2mm}
\includegraphics[width=0.495\linewidth]{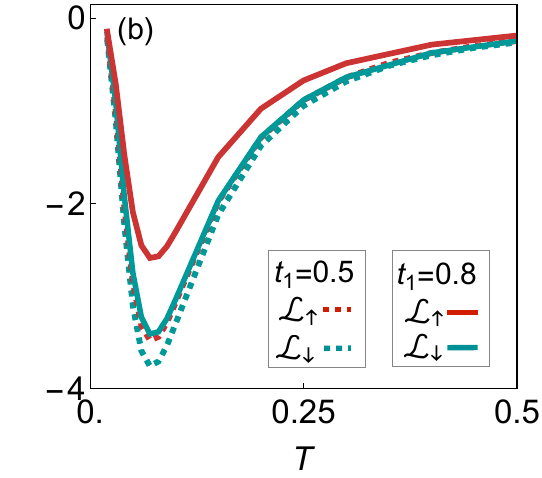}\\
\caption{
The spin-dependent thermoelectric currents $\mathcal{L}_{\sigma}$ (in units of $ek_{B}/h$) is displayed for the up-$(\mathcal{L}_{\uparrow})$ and down-spin $(\mathcal{L}_{\downarrow})$ quasiparticles with respect to (a) the AM strength $t_{1}$ and (b) the heterostructure temperature $T$ (in units of $T_{c}$) for different AM strengths $t_{1}$. This result is presented for the normal AM-SC interfaces, considering $t_{2}=0$. 
We choose temperature $T=0.3T_{c}$ for (a). The other system parameters remain the same as we consider in Fig.~\ref{L_vs_phi}.}
\label{L_vs_t:JJ}
\end{figure}

To further investigate the influence of the AM parameters on the thermoelectric current in our JJ, we analyze the behavior of $\mathcal{L}_{\sigma}$ for a wide range of $t_{1}$, keeping $t_{2}=0$, as shown in Fig.~\ref{L_vs_t:JJ}(a). Depending on the relative strength of the AM parameter to the normal part of the AM, we define two regimes: (i) $t_{1}\le t_{0}$ which essentially describe a weak AM phase and (ii) $t_{1}> t_{0}$ which is considered as the strong AM phase~\cite{Fu2025}. We find that in the weak AM phase, both the up- and down-spin currents decrease with increasing $t_{1}$, with the rates being different for the two spins. Notably, the up-spin quasiparticle current becomes zero at $t_{1}=t_{0}=1$, while the corresponding down component is finite. In contrast, in the strong AM regime, the down-spin current $\mathcal{L}_{\downarrow}$ starts increasing for further increase in $t_1$, while the other spin current continues to remain vanishingly small. These behaviors of the spin-resolved currents ensure that by tuning the AM parameter $t_1$ only, we gradually enter the strong AM phase to have quasiparticles of only one spin type contribute to the thermoelectric current, offering potential for spin-selective thermoelectric transport.

To understand the physical origin of the influence of the AM strength on the thermoelectric current, we investigate how the incident quasiparticle momenta $k_{e\uparrow}$ and $k_{e\downarrow}$ evolve with $t_{1}$. The response of these up- and down-spin momenta to $t_1$ are different, which increases significantly for large $t_1$, leading to a notable asymmetry between the spin-up and -down components of the thermoelectric current in the weak AM phase. In the strong AM regime, the spin-up quasiparticle momentum 
vanishes resulting in a complete suppression of the up-spin current. In contrast, the down-spin particle momentum survives and even grows in amplitude, and thus $\mathcal{L}_{\downarrow}$ gradually increases with $t_{1}$ in the strong phase. Note that, we also confirm that introducing a finite $t_{2}$ does not affect this behavior of the current qualitatively. However, as $t_{2}$ increases, it starts to compete with $t_{1}$ and generates a crossover between the up- and down-spin mediated thermoelectric currents. This reversal in spin dominance at larger $t_{2}$ is consistent with the behavior of the current shown in Fig.~\ref{L_vs_phi}. 

\begin{figure}
\centering
\includegraphics[width=0.5\linewidth]{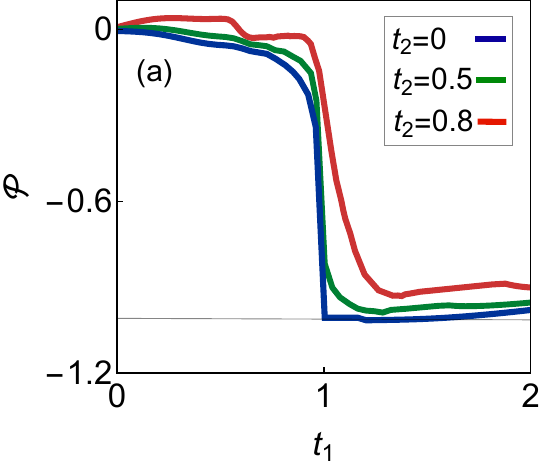}\hspace{-2mm}
\includegraphics[width=0.5\linewidth]{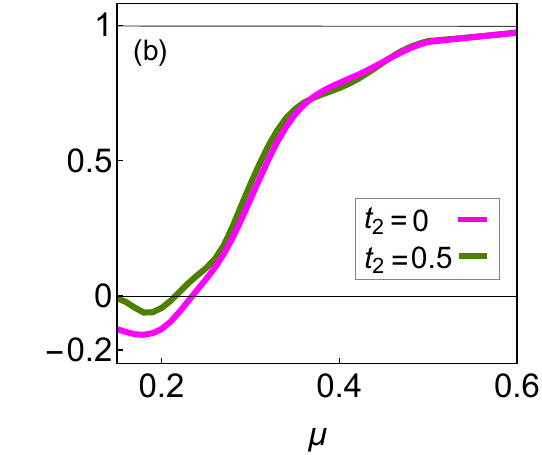}
\caption{The spin-polarization for the spin-dependent thermoelectric quasiparticle current (a) with respect to the AM strengths $t_{1}$ choosing different values of the other AM patameter $t_{2}$ and considering the chemical potential $\mu=0.2$, (b) for a range of chemical potential $\mu$, at different AM strengths of $t_{2}$ and fixed $t_{1}=0.8$. The other system parameters remain the same as we consider them in Fig.~\ref{L_vs_phi}}.
\label{P_vs_t:JJ}
\end{figure}
\subsubsection{Effect of temperature}
The behavior of the thermoelectric current in the AM-SC bilayer is already illustrated in Sec.~\ref{bilayer} for the weak AM phase [see Fig.~\ref{L_vs_t:bilayer} (b)]. We confirm that the qualitative behavior of the current remains same for the JJ too when we are in the weak AM phase. In Fig.~\ref{L_vs_t:JJ}(b), we show the variation of the thermoelectric current with respect to the base temperature of the JJ for different AM strengths $t_{1}$ considering both weak and strong regimes at $t_{2}=0$. We observe that increasing the AM strength $t_1$ enhances the disparity between the spin-resolved currents $\mathcal{L}_{\uparrow}$ and $\mathcal{L}_{\downarrow}$, which is consistent with the trends found in Fig.~\ref{L_vs_t:JJ}(a). Note that, the behavior of the current with the variation in the chemical potential is also similar to that in the bilayer junction. We also confirm that the difference between the two spin-resolved thermoelectric currents diminishes when $\mathcal{L}_{\sigma}$ is studied by varying $t_{2}$ keeping $t_{1}$ fixed (results not shown here). This behavior along with its implications for the spin polarization are further illustrated in the following text.

\subsubsection{Spin polarization}
In this subsection, we analyze how the difference between the up- and down-spin thermoelectric currents evolves with the variation of AM strengths. We define the spin polarization as
\begin{equation}
    \mathcal{P}=\frac{\mathcal{L}_{\uparrow}-\mathcal{L}_{\downarrow}}{\mathcal{L}_{\uparrow}+\mathcal{L}_{\downarrow}}\ .
\end{equation}
In Fig.~\ref{P_vs_t:JJ}, we depict the behavior of the spin polarization as a function of the AM parameter $t_1$ and the chemical potential $\mu$. In Fig.~\ref{P_vs_t:JJ}(a), we observe that the spin polarization increases gradually with respect to the AM parameter $t_{1}$ in the weak AM phase ($t_{1}\le 1$). This behavior is expected as the AM strength enhances the spin splitting and generates a higher spin-polarized thermoelectric current.
 In the strong AM regime ($t_{1}\ge 1$), the up-spin thermoelectric current vanishes [see Fig.~\ref{L_vs_t:JJ}(a)], resulting in a fully spin-polarized behavior. Specifically, {\it a $100\%$ spin-polarized thermoelectric current is achieved} when the AM phase $t_{1}$ matches with the bare kinetic energy scale, i.e., at $t_{1}=t_{0}$ with $t_{2}=0$.

As $t_{2}$ increases, however, the perfect spin polarization is no longer maintained since the wave vector for the up spin $k_{e(h)\uparrow}$ starts growing. Remarkably, when $t_{2}$ becomes comparable or larger than $t_1$, the competition between the two AM strengths causes an effective rotation of the AM-SC interface with respect to the direction of the incident particle and results in a sign change in the spin polarization even in the weak AM phase of $t_{1}$, especially for large $t_{2}$. Such spin reversal of $\mathcal{L}_{\sigma}$ to $\mathcal{L}_{\bar{\sigma}}$ for $t_{2}\ge t_{1}$ corresponds to the switching dominance of the up- and down-spin quasiparticles  observed in Fig.~\ref{L_vs_phi}(c).

Then, we analyze whether the spin polarization depends on the chemical potential too. In Fig.~\ref{P_vs_t:JJ}(b), we display the behavior of $\mathcal{P}$ as a function of $\mu$ in the absence and presence of $t_2$ for a comparatively stronger AM strength of $t_{1}$. Interestingly, we observe a sign-changing behavior of the spin polarization by tuning the chemical potential. This highlights that the relative contributions of up- and down-spin quasiparticle currents are sensitive to the electronic structure near the Fermi level, and can be effectively controlled via a gate potential. As we tune the chemical potential, the DOS of the AM shifts away from the superconducting gap $\Delta_{0}$. Hence, though the number of quasiparticle states above the superconducting gap increases due to thermal excitation, the shift in the available energy states in the DOS effectively reduces the thermoelectric current. As a consequence, the spin polarization saturates at higher values of the chemical potential. We also find that the effect due to the rotation of the AM-SC interface, which arises by tuning the AM parameters $t_{1}$ and $t_{2}$, can also be effectively captured by tuning the chemical potential of the junction. This confirms that {\it a gate control of $\mu$ provides an alternative route to modulate the spin-dependent thermoelectric response}, enhancing the potential of our model setup from the application perspective.

\section{RSOI-altermagnet based Josephson junction: Thermoelectric diode}\label{SC_AM_SC:JD}
\begin{figure}
\centering
\includegraphics[scale=0.41]{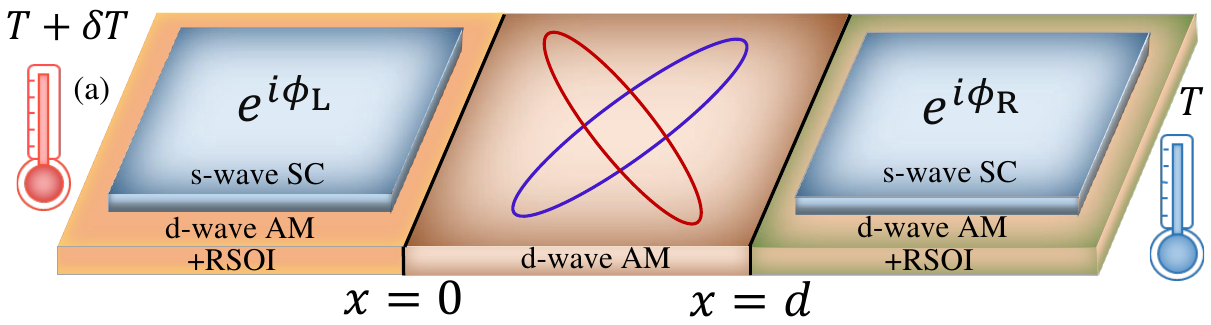}\\
\includegraphics[scale=0.45]{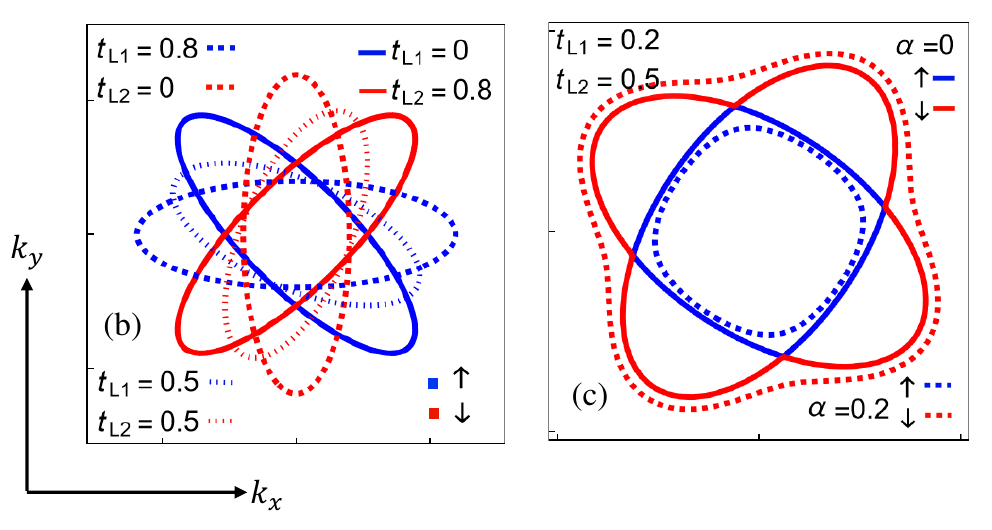}
\caption{(a) A schematic diagram of our AM-based JJ setup with RSOI $\alpha_{\text{L(R)}}=\alpha$ and proximity-induced superconductivity in the leads. 
(b) the Fermi surfaces of a bare AM (central region) in terms of the 
AM parameters $t_{\text{L1}}$ and $t_{\text{L2}}$ for various AM strengths manifesting the rotation of the AM lobes. (c) The lead AM's Fermi surface with the AM parameters $t_{\text{L1}}=0.2$ and $t_{\text{L2}}=0.5$ in the absence (solid lines) and presence (dotted lines) of RSOI ($\alpha$).}
 \label{fig:AMSC_AM_AMSC}
\end{figure}
With the understanding of the behavior of the thermoelectric current in the AM-SC bilayer and JJ, we now focus on establishing the nonreciprocity of the thermoelectric current in a thermally driven AM-based JJ with Rashba spin-orbit coupling (SOC). The origin of the diode effect in the thermoelectric current can be understood from the symmetry analysis in the heterojunction. In the literature, it has been widely demonstrated that the necessary conditions for the nonreciprocal current in a JJ are the simultaneous breaking of inversion symmetry ($\mathcal{IS}$) and time-reversal symmetry ($\mathcal{TRS}$). These broken symmetries have been utilized both in the SC phase biased~\cite{Ando2020, Misaki2021, Baumgartner2022, Zhang2022, Nagaosa2022, Davydova2022, Banerjee2024, Mao2024, Debnath2024, Debnath2025, Roy2025, Cheng2024, Sharma2025, Chakraborty2025}, and thermally biased~\cite{Chatterjee2024, Antola2024, Balduque2025} systems to explore diode effect. Additionally, recent results on Josephson diode effect in AM-based JJs have demonstrated that the presence of the crystal symmetry in the AM plays a crucial role which is beyond the established symmetry-based framework of the diode effect in AM-based junctions~\cite{Banerjee2024, Cheng2024, Sharma2025}. The majority of these works are based on Cooper pair transport due to the SC phase bias across the leads in JJ. On top of that, the control of the spin-dependent thermoelectric current in a thermally-biased AM-based JJ hints towards the possibility of generating \textit{SC phase-dependent dissipative nonreciprocal current} in the JJ, which completely remains an open question.

Consideration of the AM as a weak link within the JJ is sufficient to break the $\mathcal{TRS}$ of the junction due to the intrinsic spin splitting of the AM, as the AM violates $\mathcal{T} \mathcal{H}_{\text{AM}}(\bold{k},
\sigma) \mathcal{T}^{-1}=\mathcal{H}_{\text{AM}}(\bold{-k},\sigma)$. To break the $\mathcal{IS}$, we replace the ordinary $s$-wave SCs of the JJ by two AMs with proximitized $s$-wave superconductivity. This geometry also addresses the intriguing question: What happens if we use proximity-induced superconductivity in place of ordinary SCs? However, this AM-based JJ protects the reciprocal behavior of the thermoelectric current, i.e., the forward and backward thermoelectric currents are exactly same to each other, indicating that breaking $\mathcal{TRS}$ and $\mathcal{IS}$ are necessary but not sufficient to achieve nonreciprocal thermoelectric current in the present case. An additional symmetry breaking is necessary in AM-based JJs. From the intrinsic crystal symmetry of the AM, breaking the in-plane rotational symmetry of the AM is needed to fully break the inversion symmetry in our AM-based JJ. We consider Rashba SOC which we utilize to obtain finite thermoelectric diode effect, analogous to the Josephson diode effect in phase-biased AM-based junctions without applying any thermal bias~\cite{Cheng2024, Sharma2025, Chakraborty2025}.

To quantify the symmetry analysis, we define an in-plane $\mathcal{IS}$ operator as $\mathcal{J} \mathcal{H}(\beta_{{\text{L(R)}}}) \mathcal{J}^{-1}=\mathcal{H}(\beta_{\text{L(R)}})$, where $\beta_{\text{L(R)}}=\frac{1}{2}\text{tan}^{-1}\left(\frac{t_{\text{L2(R2)}}}{t_{\text{L1(R1)}}}\right)$ is the angle between the lobes of the left (right) AM. Despite considering different AM strengths in the two AMs, i.e., $(t_{\text{L1}},t_{\text{L2}})\ne (t_{\text{R1}},t_{\text{R2}})$, $\mathcal{IS}$ can still be protected through $\beta_{\text{L}}=\beta_{\text{R}}$. Hence, we define a new operator $\mathcal{\Tilde{J}}=\mathcal{J}\mathcal{R}$, where $\mathcal{R}$ is an operator that breaks the in-plane rotational symmetry of the lead AMs in presence of Rashba SOC leading to $\mathcal{\Tilde{J}}\mathcal{H} (\beta_{\text{L(R)}}) \mathcal{\Tilde{J}}^{-1}=\mathcal{H}(2\pi-\beta_{\text{L(R)}})$. This is described as the mirror symmetry of altermagnet~\cite{Cheng2024, Sharma2025}. These symmetry operators are simultaneously applied to the intermediate AM, which remains invariant under this operation. Therefore, though a change in the ratio of lead AMs exchange strengths is effective in principle to achieve the nonreciprocity, the presence of Rashba SOC is additionally requisite to break the crystal rotational symmetry in the lead AMs to obtain a pronounced diode effect.
\subsection{Model and method}
We consider two AMs in close proximity to ordinary SCs in the presence of Rashba SOC and connected via a normal $d$-wave AM as presented in Fig.~\ref{fig:AMSC_AM_AMSC}(a). The inclusion of the RSOI within the leads adds the spin rotational asymmetry in the junction. This drastically modifies the Fermi surfaces of the two AMs [see Fig.~\ref{fig:AMSC_AM_AMSC}(c)]. The same has been schematically illustrated in Fig.~\ref{fig:AMSC_AM_AMSC}(b) when the lobes of the AM are rotated by various choices of $t_{\text{L1}},~t_{\text{L2}}$. Hence, we write the Hamiltonian of the AM-based JJ with RSOI as 
\begin{equation}
    \mathcal{H}^{\text{JD}}=\mathcal{H}^{\text{JJ}}+\sum_{\nu\in\{\text{L,R}\}} \left(\Gamma^{\nu}_{\boldsymbol{k}} \hat{\tau}_{z}\hat{\sigma}_{z}+\hat{\tau}_{z} \lambda^{\nu}_{\boldsymbol{k, so}}\right)\ ,
    \label{ham:JD}
\end{equation}
where, $H^{\text{JJ}}$ is the same as mentioned in Eq.~(\ref{ham:JJ}). The additional terms corresponding to the leads are taken as,
\begin{eqnarray}
   \Gamma^{\nu}_{\boldsymbol{k}}& = &[t_{\nu 1} (k_{y}^{2}-k_{x}^{2})+2t_{\nu 2} k_{x}k_{y}]\Theta(-x) \Theta(x-d)\ ,\label{Ham:AM_JD} \nonumber\\
   \lambda^{\nu}_{\boldsymbol{k, so}}& = & \alpha_{\nu}(\sigma_{y} k_{x}-\sigma_{x} k_{y}) \Theta(-x) \Theta(x-d)\ ,
\end{eqnarray}
where $t_{\nu i}$ with $\nu \in \{\rm {L,R} \}$ and  $i\in \{1,2\}$ correspond to the parameter for the AM leads and $\alpha_{\nu}$ is the RSOI strength at $\nu$th proximity-induced AM lead. For the sake of simplicity, we consider $\alpha_{\text{L}}=\alpha_{\text{R}}=\alpha$ which does not affect 
our results qualitatively. 
\begin{figure}
\vspace{-2mm}
\includegraphics[scale=0.474]{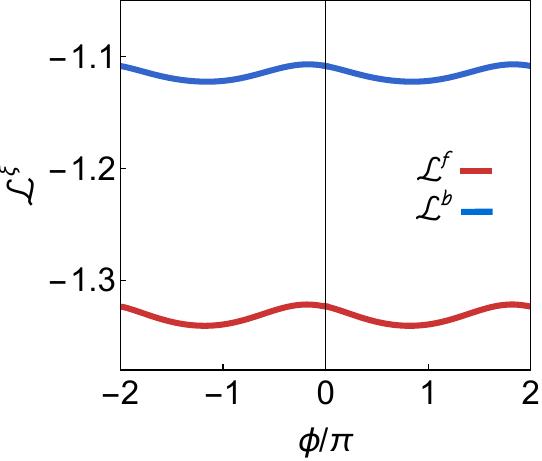}
\includegraphics[scale=0.45]{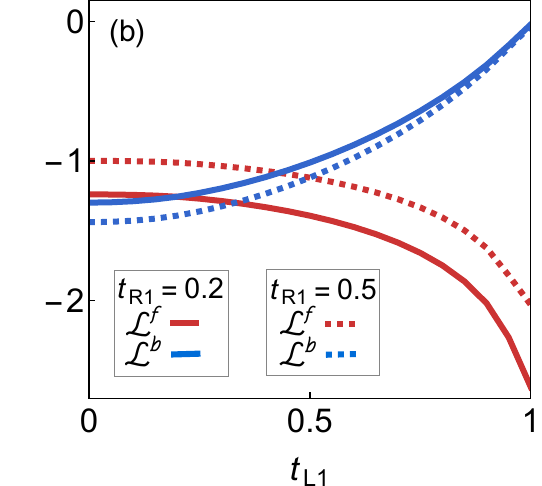}
\caption{(a) The forward ($\mathcal{L}^{f}$) and backward ($\mathcal{L}^{b}$) thermoelectric currents ($\mathcal{L}^{\zeta}$) (in units of $ek_{B}/h$) with respect to the superconducting phase difference $\phi$ with the strength of the middle AM as $t_{1}=0.8$ and $t_{2}=0$. The model parameters 
for the leads are considered as $t_{\text{L1}} = 0.4, t_{\text{L2}} = 0.3, t_{\text{R1}} = 0.2, t_{\text{R2}} = 0.1$, and the RSOI strength is considered as $\alpha=0.2$. The other system parameters are considered to be the same as mentioned in Fig.~\ref{L_vs_phi}. (b) The forward and reverse currents (in units of $h/ek_{B}$) with respect to the strength of the left AM lead $t_{\text{L1}}$ for various $t_{\text{R1}}$ in the right lead and $\phi=\pi/4$. The other model parameters remain same as mentioned in panel (a).}
\label{L_vs_phi:JD}
\end{figure}
We use Eq.~(\ref{ham:JD}) to solve the Schr\"{o}dinger equation using the modified boundary conditions (see Appendix~\ref{scatmat:JD} for details). The scattering coefficients found in this way are utilized to investigate the thermoelectric current $\mathcal{L}^{\zeta}$ with $\zeta \in \{f, b\}$ to denote the forward and backward current in our RSOI-AM-based JJ. We define the forward current $\mathcal{L}^{f}$ carried by quasiparticles from the left to the right lead, whereas the backward current is defined as the current flows from the right to the left lead as $\mathcal{L}^{b}$. The diode effect is found when $\mathcal{L}^{f} \ne  \mathcal{L}^{b}$. We study the diode effect for various parameter regimes in the following subsections.


\subsection{Results and discussions}
In this subsection, we present our results for the thermoelectric current in the RSOI-AM-based JJ and discuss the diode effect in the thermoelectric current 
as a function of the superconducting phase difference, AM strength, junction temperature, and chemical potential.
\subsubsection{SC phase tunability and effect AM parameters}
Siimilar to the previous section, we analyze the tunability of the thermoelectric current by the SC phase difference and the influence of the AM parameters in the presence of Rashba SOC. In this regard, we demonstrate the behavior of the forward and backward thermoelectric currents as a function of the SC phase difference $\phi$ in Fig.~\ref{L_vs_phi:JD}(a). Note that, throughout the present section, we consider only the total current instead of the spin-resolved current to focus on the direction-dependent thermoelectric current since spin is not a good quantum number here due to the presence of the spin-mixing term induced by the Rashba SOC. We observe that the behavior of both the forward and backward currents are oscillatory which is similar to what we found in case of AM-JJ without Rashba SOC as discussed in the previous section. We consider a combination of the AM parameter strengths $\{t_{\text{L}1}, t_{\text{L}2}\}$ and $\{t_{\text{R}1}, t_{\text{R}2}\}$, which present different crystal plane symmetries of the AM lobes and, thus, it causes distinct differences between the 
forward and backward currents in the JJ.

To understand the difference between the two currents and how it evolves with the parameter values in more detail, we depict the direction-dependent thermoelectric current as a function of the AM strength $t_{\text{L}1}$ of the left lead for various AM parameters of the right lead $t_{\text{R}1}$, in Fig.~\ref{L_vs_phi:JD}(b). We find that the amplitude of the forward current increases and that of the backward current reduces with increasing $t_{\text{L1}}$. It is valid for all values of $t_{\text{R}1}$. Since the variation of $t_{\text{L1}}$ effectively rotates the AM-SC interface in the left lead, the forward current increases with the change in $\beta_{\text{L}}$, though the reverse current due to the incoming particles from the right lead hardly traces the changes occurring in the left lead, that effectively reduces the backward current with $t_{\text{L}1}$. These variations of the forward and backward currents are independent of $t_{\text{R1}}$, though the difference between the forward and backward currents is modified with $t_{\text{R}1}$. Interestingly, we find a symmetric point at $t_{\text{R}1}=t_{\text{L}1}=t_{c}$ where the forward and backward currents cross each other indicating exactly the same values for both currents, i.e., vanishing diode effect. This crossing point of the forward and reverse currents implies that this specific value of $t_{\text{L}1}$ acts as a critical point ($t_{c}$) for the nonreciprocity in the junction.

\begin{figure}
\centering
\includegraphics[width=4.25cm]{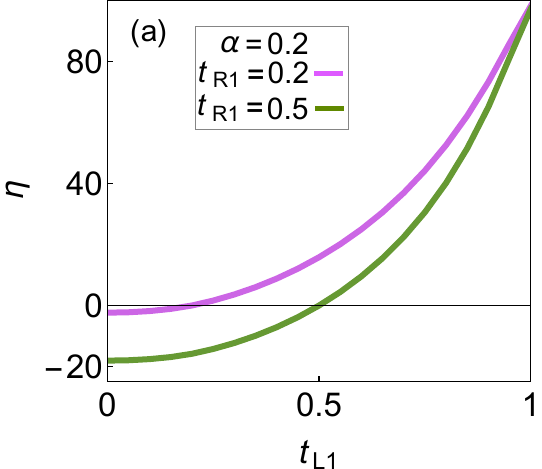}
\includegraphics[width=4.25cm]{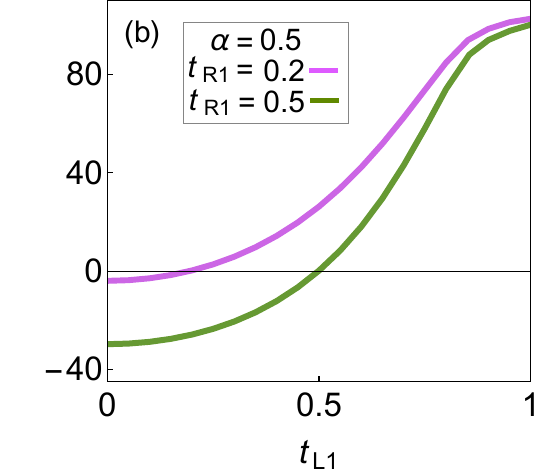}
\caption{The diode efficiency $\eta$ as a function of the left lead's AM strength $t_{\text{L1}}$ for different Rashba SOC strengths $\alpha$ in the lead AM. The weaklink middle AM's parameters are considered as $t_{1}=0.8$ and $t_{2}=0$. The other AM parameter strengths of the leads are $t_{\text{L2}} = 0.2,~t_{\text{R2}} = 0.1$. The other model 
parameters are considered to be the same as mentioned in Fig.~\ref{L_vs_phi}.} 
\label{fig:JD}
\end{figure}
\subsubsection{Thermoelectric diode efficiency}
To quantify the difference between the forward and backward currents, we define the diode efficiency of the thermoelectric current as
\begin{eqnarray}
\eta=\left(\frac{\mathcal{L}^{f}-\mathcal{L}^{b}}{\mathcal{L}^{f}+\mathcal{L}^{b}}\right) \times 100\ \% \ ,
\end{eqnarray}
where, $\mathcal{L}^{f}$ and $\mathcal{L}^{b}$ have the same meaning as described earlier. We show the behavior of the diode efficiency as a function of the 
left AM parameter $t_{\text{L}1}$ for various values of the other parameters of the AM in Fig.~\ref{fig:JD}. In Fig.~\ref{fig:JD}(a), we observe that the diode efficiency increases monotonically as one increases $t_{\text{L1}}$, but it reduces with increasing $t_{\text{R1}}$. This indicates that by rotation of the 
right lead AM-SC interface, the backward current increases, and thus it effectively reduces the diode efficiency. Remarkably, the efficiency changes its sign when $t_{\text{L1}} = t_{\text{R1}}$, which is consistent with our observation of the critical point for the thermoelectric current in Fig.~\ref{L_vs_phi:JD}(b). 
This also emphasizes that by tuning the angle of the AM-SC interface by the rotation of the lobes of AM leads, the forward and reverse currents can be controlled and the direction of the current can be tuned as per preference.

In the present model, the Rashba SOC plays a crucial role. To analyze how the diode effect changes with the Rashba SOC strength, we illustrate the diode efficiency in Figs.~\ref{fig:JD}(a) and~\ref{fig:JD}(b) for two different Rashba SOC strengths. Comparing Figs.~\ref{fig:JD}(a) and~\ref{fig:JD}(b), we see that the diode efficiency increases with the increase of the RSOI strength. Specifically, in the negative efficiency regime where the backward current dominates over the forward current, the diode efficiency becomes almost the double in the weak AM phase. In contrast, for the strong AM phase, although the magnitude does not change significantly, the higher values of the diode efficiency are found for wider ranges of $t_{\text{L1}}$ compared to the case of weak RSOI. Overall, the diode efficiency is elevated for the stronger RSOI.

\section{Summary and Conclusion}\label{Summary}
To summarize, in this article, we have systematically investigated various AM-based heterojunctions to explore the thermoelectric current carried via quasiparticles in external field-free condition. The recent advent of AM materials, like $\text{RuO}_{2}$, $\text{MnTe}$, $\text{CrSb}$, etc., have been shown to manifest finite energy band splitting due to the lifted Kramer's spin degeneracy~\cite{Krempasky2024}. We explore the effect of this AM spin splitting on the thermoelectric phenomena in AM-based superconducting junctions. Our study focuses on the low-energy quasiparticle excitations, ignoring the phonon contributions to the thermoelectric current, which is justified in the low-temperature regime. Within the linear response regime, based on the Onsager's theory in quantum transport, we calculate the thermoelectric quasiparticle current generated in the AM-based SC junctions driven by the temperature gradient.

Starting with the minimal bilayer model of AM-SC heterojunction, we find that the AM parameter with $d_{x^{2}-y^{2}}$ symmetry is the primary driver for the spin-split thermoelectric current, which can be enhanced by tuning the AM parameter $t_{1}$. The spin-dependent thermoelectric quasiparticle current flows in the AM-JJ too and, interestingly, in the strong AM phase, the difference between the up- and down-spin currents results in $100\%$ spin polarization of the thermoelectricity. We analyze the effects of junction temperature and chemical potential for a wide range of AM system parameters. We also confirm that the quasiparticle mediated thermoelectric current in our thermally biased AM-JJ arises predominantly due to the dissipative component, with negligible nondissipative current component, which makes our AM-based junctions as an efficient testbed for spin-caloritronics.

We further study the AM-based JJ including RSOI in search of nonreciprocity in the behavior of the thermoelectric current. Note that, here superconductivity has been assumed to be induced in the AM via the proximity effect. We explore the effects of the in-plane rotational symmetry of the AM that protects the nonreciprocity of the thermoelectric current. Therefore, to establish the thermoelectric diode effect in the AM-based JJ, we consider altermagnetic SC with RSOI as the leads, and this justifies the symmetry requirements for the diode effect in our AM-based JJ. Additionally, varying the AM strengths in the leads, which gives rise to the rotation of the AM lobes and effectively modifies the AM-SC interface at the junction, and by considering different RSOI strengths, we investigate how these parameters influence the  nonreciprocity of the thermoelectric current.

The experimental realization of the AM-SC heterojunction may be possible to achieve with the choice of possible $d$-wave AM, like $\text{RuO}_{2}$ or $\text{MnO}_{2}$ etc., and forming a junction with ordinary SCs like Nb, Al, or Pb etc., which typically exhibits a transition temperature $T_{c}\sim 2-10$K and the SC gap is typically of the order of meV, as an example $2\Delta \sim 3.1$ meV in Nb. To introduce nonreciprocity in the altermagnetic SC JJ, RSOI can be introduced to the lead AMs by an external electric field or by controlling the effective potential of the junction of the heterostructure by an external gate voltage or by introducting a layer with RSOC. On the other hand, rotation of the AM angle $\theta$ can be conrolled by modifying the parameters $t_{1}$ and $t_{2}$ via strain engineering~\cite{Belashchenko2025, Chakraborty2024}, which effectively modifies the lattice parameters of the material. For example, $\text{OsO}_2$ is known to be a $d$-wave altermagnet and a recent strain engineering on this lattice has achieved the deformation of lattice parametter of about $2\%\sim 5\%$ which effectively modifies the Fermi surface of the altermagnet~\cite{Zhang2025}. This replicates the effective theoretical modulation of the $t_{1}$ and $t_{2}$ parameters in our study. Our results propose the achievement of nearly $100\%$ nonreciprocity in the thermoelectric particle current in the weak AM regime with enhanced experimental precision.

In conclusion, thermoelectricity offers an efficient way for technology based on energy-saving device components, and the recent advent of unconventional AMs add 
very promisizng prospects to that. Our study supports this prospect in positive aspect. Finally, a few comments on this are in order. The recent inception of three-dimensional AMs~\cite{Jeschke2024, Yang2025} leaves a greater opportunity to explore the thermal effects in such heterostructures. Additionally, as a few studies also indicate the existence of unconventional magnetic materials~\cite{Liu2025, Fukaya2025} as well as unconventional SC have shown a few insights into thermoelectric properties, hence it could be a great deal for future exploration of thermoelectricity arising from these \textit{unconventionals}.
\acknowledgments{D.\,D. and P.\,D. acknowledge the Department of Space, Government of India, for all support at Physical Research Laboratory (PRL). D.\,D. acknowledges M. Papaj for useful communications and P. Chatterjee for some initial discussions. D.\,D. also thanks L. \v{S}mejkal and W. Campos for mentioning relevant references.}
 
\begin{appendix}
\appendix
\section{Density of states (DOS) for AM}\label{dos}
In this appendix, we show the DOS of the bulk $d$-wave AM to explore the effect of AM parameters on the intrinsic spin splitting. The spin-resolved DOS of the AM is defined as
\begin{equation}
    \rho_{\sigma}(E)=-\frac{1}{\pi} \int \text{Tr}[\text{Im} [G_{\sigma}(E)] ] dk_{x} dk_{y}\ ,
    \label{eq:dos}
\end{equation}
where, $G_{\sigma}(E)$ is the Green's function of the AM defined as $G_{\sigma}=(E-H_{\sigma})^{-1}$. Here, $H_{\sigma}$ is the corresponding spin component of the Hamiltonian which can be found after diagonalizing the Hamiltonian and the local DOS is obtained by integrating over the $(k_x,k_y)$ plane as $k_{x}\in [0,2\pi]$ and $k_{y}\in[0,k_{y,\text{max}}]$, where $k_{y,\text{max}}$ [see Eqs.~\eqref{ky_max_e} and ~\eqref{ky_max_h}] is the highest extent of electron (hole) energy contour (hyperbolic surface) of a given spin which limit the extent of Andreev reflection with the opposite spin (as explained in detail in Refs.~\cite{Papaj2023, Sun2023}). This limit of $k_{y,\text{max}}$ leads to the spin splitting in the DOS, whereas the net magnetization over the entire energy range remains zero, satisfying the characteristic properties of the AM. Note that, spin-resolved components can be written in this way as long as there is no spin mixing in the Hamiltonian and spin is a well-defined quantum number. The total DOS of the AM can be found by considering the total contributions of the two spin components as $\rho(E)=\sum\limits_{\sigma}\rho_{\sigma}(E)$. It is important to mention that the DOS becomes identical for the up- and down-spin quasiparticles when the integration is carried over the full Brillouin zone as $(k_x,k_y)\in [-\pi,\pi]$.
\begin{figure}
\centering
\hspace{-2mm}
\includegraphics[width=0.48\linewidth]{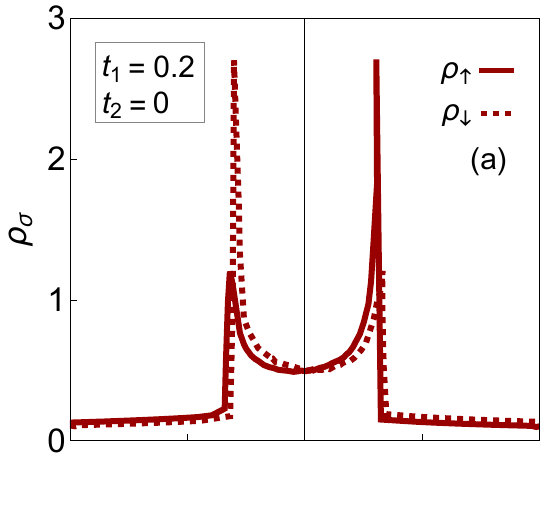} \hspace{0.2mm}
\includegraphics[width=0.46\linewidth]{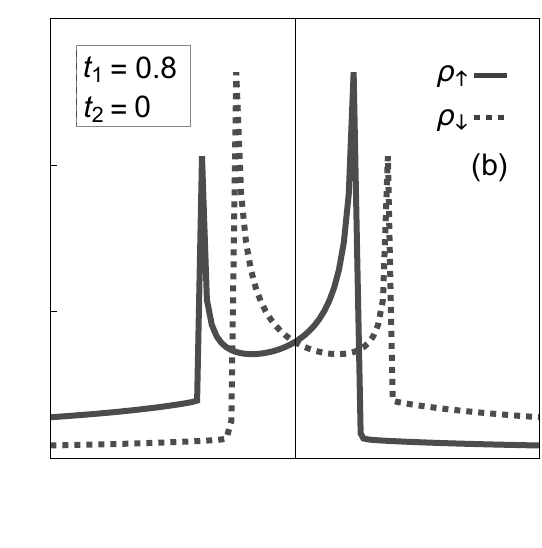}\\
\vspace{-4mm} \hspace{-2mm}
\includegraphics[width=0.48\linewidth]{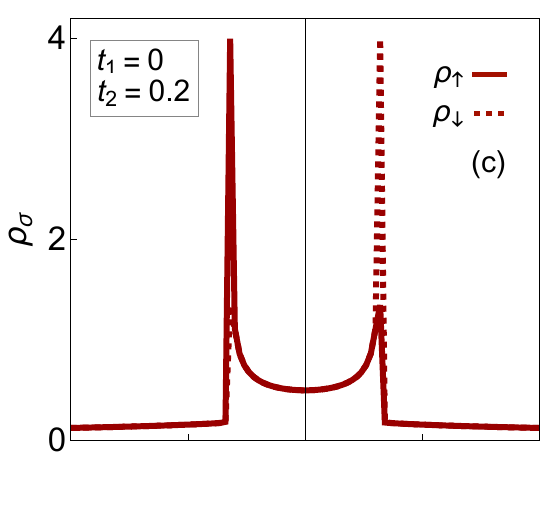}  \hspace{0.1mm}
\includegraphics[width=0.46\linewidth]{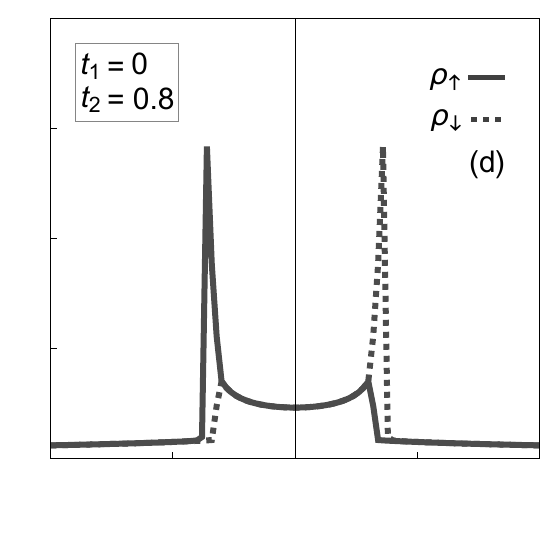} \\
\vspace{-4mm} \hspace{0.1mm}
\includegraphics[width=0.49\linewidth]{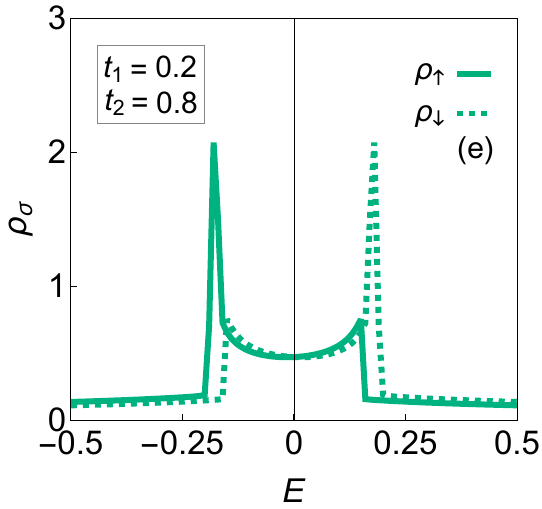} 
\includegraphics[width=0.475\linewidth]{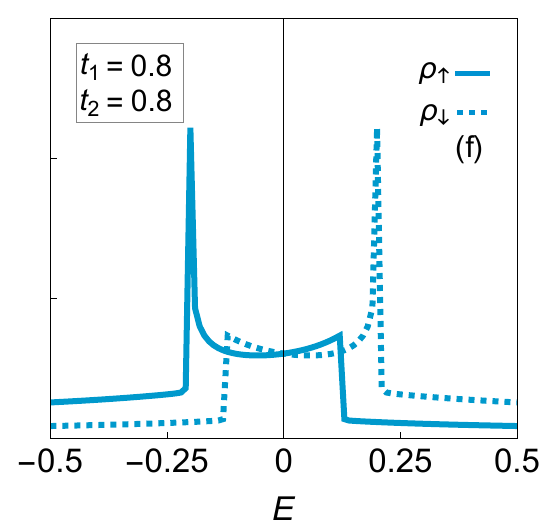}
\caption{The features of DOS, corresponding to a bulk AM, are depicted with respect to the energy of the quasiparticle in AM for different sets of AM parameters $t_{1}$ and $t_{2}$. We consider here $\mu=200\Delta_{0}$. The choice of $t_{1}$ and $t_{2}$ suggest that (a) and (b) represent the AM-SC interface at normal incidence for the incoming particle. (c) and (d) are for the $45^{o}$ rotated AM-SC interface, and (e) and (f) present the arbitrarily rotated AM-SC interface for the incident particle.}
\label{fig:dos}
\end{figure}
We study the effects of the AM parameters on the spin-split 
DOS profiles explicitly and present in Fig.~\ref{fig:dos}. In Fig.~\ref{fig:dos}(a), we find that the spin splitting of the $d_{x^2-y^2}$ symmetric AM appears due to the presence of finite $t_{1}$ by setting the other AM parameter as $t_{2}=0$. The degree of spin splitting increases with the enhancement in the value of $t_1$ as shown in Fig.~\ref{fig:dos}(b). It results in the appearance of one of the spin DOS finite at some particular energy values while the DOS for the other spin component is zero. The exact opposite scenario is obtained for the opposite energy value.

Then, we turn on $t_{2}$ while $t_{1}$ is switched off. In the presence of a finite $t_{2}$, we see in Fig.~\ref{fig:dos}(c) that there is no explicit spin splitting except the asymmetry 
in the spin-dependent DOS. By increasing $t_{2}$ in Fig.~\ref{fig:dos}(d), we find that the DOS is marginally modified, with the absence of spin splitting. To find the simultaneous effects of both the AM parameters on the DOS, we refer to Figs.~\ref{fig:dos}(e)-~\ref{fig:dos}(f). 
Presence of both finite $t_{1}$ and $t_{2}$ simultaneously, convey a rotated AM-SC interface. Hence, by comparing Figs.~\ref{fig:dos}(a) and~\ref{fig:dos}(e), we find that the spin-split effect is modified at finite $t_{2}$ and the peak height is reduced. For the specific spin component at a constant energy, the peak height is dominated by $t_{2}$. In Fig.~\ref{fig:dos}(f), we increase $t_{1}$ for the same value of $t_{2}$ as mentioned in Fig.~\ref{fig:dos}(e). For strong $t_{1}$, the explicit difference in the up- and down-spin peaks becomes prominent even in presence of finite $t_{2}$.

Therefore, by studying the behavior of DOS of the bulk AM, we conclude that the spin splitting of an AM is almost solely controlled by $t_{1}$ wheras the shifting of peaks along the energy scale is controlled by $t_{2}$. We also find that with the change in the chemical potential $\mu$, the DOS moves away from the SC gap $\Delta$ (not shown here).

\section{Onsagar's relation}\label{onsagar}
The charge and heat currents due to the voltage bias $\Delta V$ and the temperature gradient $\Delta T$ are connected through the Onsager relation as follows:
\begin{equation}
    \left[\begin{array}{c}
\mathcal{I}_{C}  \\
\mathcal{I}_{Q}
\end{array}\right]=\left[\begin{array}{cc}
\mathcal{L}_e  & \mathcal{L}\\
\mathcal{L} & \mathcal{L}_{th}
\end{array}\right] \left[\begin{array}{c}
\Delta V \\
\Delta T
\end{array}\right]\ ,
\end{equation}
where $\mathcal{I}_{C}$ and $\mathcal{I}_{Q}$ describe the charge and heat currents, respectively. The charge conductance due to the voltage bias and the thermal conductance due to the temperature bias are defined through $\mathcal{L}_{e}$ and $\mathcal{L}_{th}$, respectively. The charge current that arises due to a finite temperature bias is known as the thermoelectric current, which is represented by $\mathcal{L}$, and the reverse process, i.e., the heat current generated due to the voltage bias known as the Peltier effect. The latter is also denoted by $\mathcal{L}$. The spin splitting of the AM can separate the spin-dependent thermoelectric currents without any external magnetic field and the total thermoelectric particle current can be denoted as $\mathcal{L}=\sum\limits_{\sigma}\mathcal{L}_{\sigma}$. Here, $\mathcal{L}_{\sigma}$ represents the spin-resolved thermoelectric current. Note that, 
due to the presence of Rashba SOC in AM-JJ, the possibility of spin flipping restricts the spin-resolved currents.

\section{Scattering matrix formalism for AM-SC bilayer}\label{scatmat}
In this appendix, we discuss some steps of our calculations based on the scattering matrix formalism implemented for the AM-SC junction in support of the discussions presented in the main text.

\subsection{Formulation}\label{scatmat:JJ1}
Due to the absence of any spin-mixing term in the Hamiltonian of the AM-SC hybrid setup [see Eq.~(\ref{ham:bilayer})] where we decouple the BdG Hamiltonian into two blocks in the reduced basis $(\psi_{k\uparrow},\psi_{-k\downarrow})^{\text{T}}$~\cite{Sun2023, Chen2025}. For the quasiparticles incident from the AM region to the interface, the wavefunction for the AM can be written by considering the flow along the $x$ axis and preserving the momentum along the $y$ axis as
\begin{widetext}
\begin{equation} 
\psi_{\text{AM}}^{e(h)\sigma}=\left(\begin{array}{c}
1  \\
0
\end{array}\right) e^{i k_{e(h)\sigma} x}+r_{\text{N}_{e(h)}}^{\sigma}\left(\begin{array}{c}
1  \\
0
\end{array}\right) e^{-i k_{e(h)\sigma} x}+
r_{\text{A}_{h(e)}}^{\sigma}\left(\begin{array}{c}
0  \\
1
\end{array}\right) e^{i k_{h(e)\Bar{\sigma}} x},
\end{equation}
\end{widetext}
where an incoming electron (hole) 
with spin $\sigma$ is incident on the interface ($x=0$) and undergoes ordinary reflection with probability of $|r_{\text{N}_{e(h)}}^{\sigma}|^{2}$ or Andreev reflection with probability $|r_{\text{A}_{h(e)}}^{\sigma}|^{2}$. We can write the wave function for the SC region as~\cite{Sun2023, Chen2025}
\begin{equation}
    \psi_{\text{SC}}^{e(h)\sigma}=t_{e(h)}^{\sigma} \left(\begin{array}{c}
\sigma u  \\
v
\end{array}\right) e^{i k^{sc}_{e(h)} x}+ t_{h(e)}^{\sigma} \left(\begin{array}{c}
\sigma v  \\
 u
\end{array}\right) e^{-i k^{sc}_{h(e)} x}\ ,
\end{equation}
where $|t_{e(h)}^{\sigma}|^{2}$ and $|t_{h(e)}^{\sigma}|^{2}$ denote the spin-resolved 
transmission probabilities for the electronlike and holelike quasiparticles, respectively, and $u(v)=\sqrt{\frac{1}{2}\left(1\pm\sqrt{1-\frac{\Delta^{2}}{E^{2}}}\right)}$.

We find the scattering coefficients using the boundary conditions for the AM-SC bilayer~\cite{Papaj2023, Sun2023} as
\begin{equation}
\psi_{\text{AM}}^{\sigma}|_{x=0}=\psi_{\text{SC}}^{\sigma}|_{x=0}\ ,
\label{bc1}
\end{equation}
\vspace{-0.5cm}
\begin{eqnarray}
    \left(t_{0} \tau_{0}-\sigma t_{1} \tau_{z}\right) \partial_{x}\psi_{\text{AM}}^{\sigma}|_{x=0^{-}}-t_{sc} \tau_{0} \partial_{x} \psi_{\text{AM}}^{\sigma}|_{x=0^{-}} 
    \nonumber \\
    =-i t_{2} k_{y} \sigma\tau_{z}\psi_{\text{SC}}^{\sigma}|_{x=0^{+}}\ .
    \label{bc2}
\end{eqnarray}
The momenta of the quasiparticles in the AM and SC regions can be calculated as
\begin{eqnarray}
     k_{e(h)\sigma}&=&\frac{\sqrt{(t_{0}-t_{1}\sigma)(\mu\pm E)-(t_{0}^{2}-t_{1}^{2}-t_{2}^{2})k_{y}^{2}}-t_{2}k_{y}\sigma}{(t_{0}-t_{1}\sigma)}\ ,\nonumber     \\
      k^{sc}_{e(h)}&=&\sqrt{\frac{\mu\pm\sqrt{E^{2}-\Delta_{\text{L}}^{2}}}{t_{\text{SC}}}-k_{y}^{2}}\ .
      \label{am_k}
\end{eqnarray}
The momenta of quasiparticles in the $y$ direction are restricted within the hyperbolic Fermi surface of the AM. When an electron (hole) is incident at the interface of the AM-SC junction, the maximum Andreev reflection probability as a hole (electron) with opposite spin is limited by the cutoff 
over $k_{y}$ defined as~\cite{Papaj2023}
\begin{eqnarray}
    k_{y,\text{max}}^{e,\sigma}&=& \sqrt{\frac{(t_{0}\mp t_{1})(\mu+E)}{t_{0}^{2}-t_{1}^{2}-t_{2}^{2}}}
    \label{ky_max_e}\ ,\\
    k_{y,\text{max}}^{h,\sigma}&=& \sqrt{\frac{(t_{0}\mp t_{1})(\mu-E)}{t_{0}^{2}-t_{1}^{2}-t_{2}^{2}}}\ .
    \label{ky_max_h}
\end{eqnarray}
Exceeding this $k_{y,\text{max}}$, all the quasiparticles encounter normal reflection.
\subsection{Role of barrier potential}\label{appen:amsc_Z}
In the presence of a tunneling barrier at the interface of the AM-SC junction [$Z\delta(x)$] with the potential strength $Z$, the second boundary condition of Eq.~\eqref{bc2} modifies as follows~\cite{Papaj2023, Sun2023, Lu2024, Chen2025}:
\begin{eqnarray}
    \left(t_{0} \tau_{0}-\sigma t_{1} \tau_{z}\right) \partial_{x}\psi_{\text{AM}}^{\sigma}|_{x=0^{-}}-t_{sc} \tau_{0} \partial_{x} \psi_{\text{AM}}^{\sigma}|_{x=0^{-}} 
    \nonumber \\
    =(-i t_{2} k_{y} \sigma\tau_{z}+2Z)\psi_{\text{SC}}^{\sigma}|_{x=0^{+}}\ .
    \label{c3}
\end{eqnarray}

\begin{figure}
    \centering
    \includegraphics[width=0.7\linewidth]{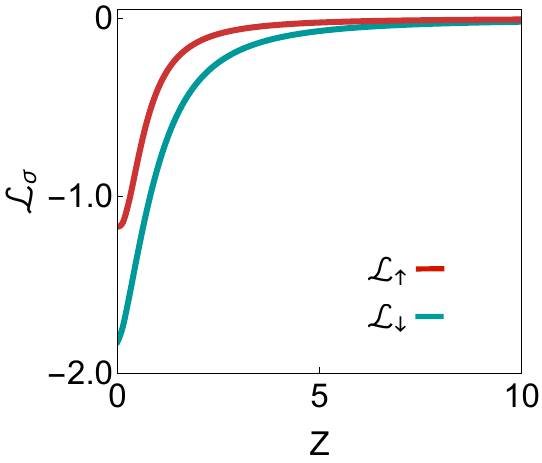}
    \caption{The thermoelectric current $\mathcal{L}_{\sigma}$ (in units of $e k_{B}/h$) as a function of interfacial tunneling barrier strength $Z$. Here, we choose $t_{1}=0.8$ and all other model parameters remain the same as mentioned in Fig.~\ref{L_vs_t:bilayer}(a).}
    \label{L_vs_z}
\end{figure}

We solve this equation in addition to the continuity of wave function [Eq.~\eqref{bc1}] and calculate the thermoelectric current using Eq.~\eqref{L_bilayer}. We find that the current is reduced with the increase in the barrier potential strength, as shown in Fig.~\ref{L_vs_z}. As the strength of $Z$ is enhanced, the probability of quasiparticle scattering via the ordinary reflection increases which essentially causes a reduction in the total current. Hence, the current gradually saturates to vanishingly small value when $Z\gg1$.
\vspace{2.5cm}
\section{Scattering matrix formalism and some other results for AM-JJ} \label{AM-JJ appendix}
In this appendix, we present the scattering matrix approach for the AM-based JJ and also discuss some of the qualitative and quantitative behavior of the total thermoelectric quasiparticle current in the AM-JJ.
\vspace{+0.4cm}
\begin{figure*}
\centering
\includegraphics[width=4.8cm]{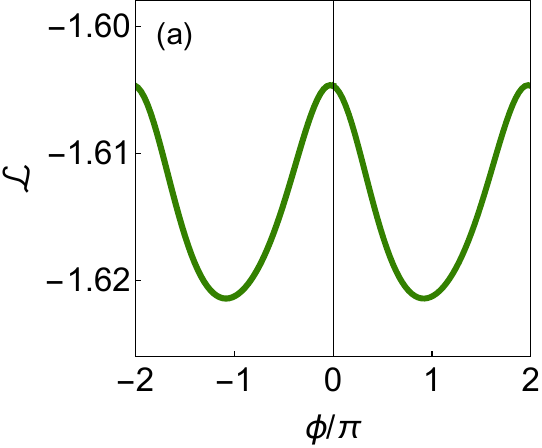}\hspace{0.25cm}
\includegraphics[width=4.8cm]{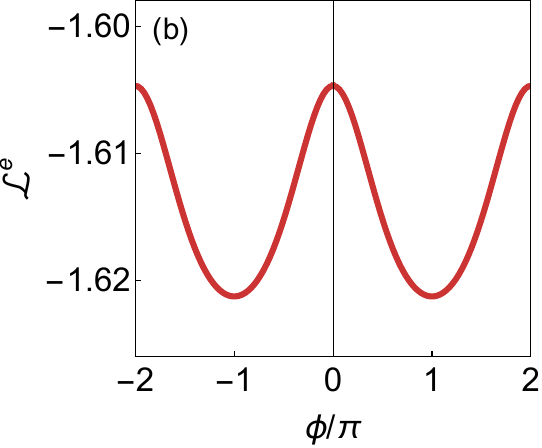}\hspace{0.25cm}
\includegraphics[width=5cm]{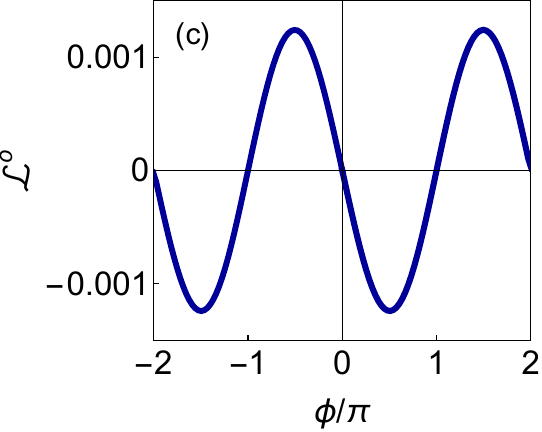}
\caption{(a) The total thermoelectric current $\mathcal{L}$ (in units of $ek_{B}/h$) with respect to the SC phase $\phi$. (b) the even-in $\phi$ ($\mathcal{L}^{e}$) and (c) the odd-in $\phi$ ($\mathcal{L}^{o}$) fraction of the total thermoelectric current $\mathcal{L}$, respectively. The other model parameters are considered to be the same as mentioned in Fig.~\ref{L_vs_phi}(b).}
\label{L_evenodd_0.8_0.5}
\end{figure*}

\subsection{Scattering matrix formulation}\label{scatmat:JJ}
Following the decoupled particle-spin basis as considered in the case of AM-SC~\cite{Sun2023, Chen2025}, the wave function for each component of the JJ is considered as
\begin{widetext}
\begin{eqnarray}
\psi_{e(h)\text{L}}^{\sigma}&=&\psi_{e\text{L}(h\text{L}),\sigma}^{+}+r_{\text{N}_{e(h)}}^{\sigma} \psi_{e\text{L}(h\text{L}),\sigma}^{-}+r_{\text{A}_{h(e)}}^{\sigma} \psi_{h\text{L}(e\text{L}),\Bar{\sigma}}^{-}\ , \\
 \psi_{\text{AM}}^{\sigma}&=&a^{\sigma} \psi_{e(h),\sigma}^{+}+b^{\sigma} \psi_{e(h),\sigma}^{-}+c^{\sigma} \psi_{h(e),\Bar{\sigma}}^{+} + d^{\sigma} \psi_{h(e),\Bar{\sigma}}^{-}\ ,\\
 \psi_{e(h)\text{R}}^{\sigma}&=&t_{ee(hh)}^{\sigma} \psi_{e\text{R}(h\text{R}),\sigma}^{+}+t_{he(eh)}^{\sigma} \psi_{h\text{R}(e\text{R}),\Bar{\sigma}}^{+}\ ,
\end{eqnarray}
\end{widetext}
where,  $\sigma$ can take $\pm 1$ values depending on the up- (down-) spin component and the wave functions are given by
\begin{eqnarray}
\psi_{e\text{L}\sigma}^{\pm}&=&\left(\begin{array}{c}
\sigma u e^{i\phi_{\text{L}}}  \\
v 
\end{array}\right) e^{\pm ik_{e\text{L}} x+ik_{y} y},\\
\psi_{h\text{L}\Bar{\sigma}}^{\pm}&=&\left(\begin{array}{c}
\sigma v e^{i\phi_{\text{L}}}\ ,   \\
u
\end{array}\right) e^{\mp ik_{h\text{L}} x+ik_{y} y}\ ,\\
\psi_{e(h),\sigma}^{\pm}&=&\left(\begin{array}{c}
1  \\
0 
\end{array}\right) e^{\pm(\mp)ik_{e(h)\sigma} x+ik_{y} y}\ ,\\
\psi_{e(h),\Bar{\sigma}}^{\pm}&=&\left(\begin{array}{c}
0\\
1 
\end{array}\right) e^{\pm(\mp)ik_{e(h)\Bar{\sigma}} x+ik_{y} y}\ ,\\
\psi_{e\text{R}\sigma}^{\pm}&=&\left(\begin{array}{c}
\sigma u e^{i\phi_{\text{R}}}  \\
v 
\end{array}\right) e^{\pm ik_{eR} x+ik_{y} y},\\
\psi_{h\text{R}\Bar{\sigma}}^{\pm}&=&\left(\begin{array}{c}
\sigma v e^{i\phi_{\text{R}}}  \\
u
\end{array}\right) e^{\mp ik_{h\text{R}} x+ik_{y} y}\ .
 \end{eqnarray}
where $k_{e\text{L}(e\text{R})}$ corresponds to the momentum of the left (right) lead quasiparticle, and $k_{e(h)\sigma}$ indicates the momentum of the electron (hole) for the intermediate AM region. 
The momenta of the AM region remain same as mentioned in Eq.~(\ref{am_k}), and  for the left and right SC leads, the momenta are the following:
\begin{eqnarray}
    k_{e\text{L}(h\text{L})}&=&\sqrt{\frac{\mu\pm\sqrt{E^{2}-\Delta_{\text{L}}^{2}}}{t_{\text{SC}}}-k_{y}^{2}}\ ,\\
     k_{e\text{R}(h\text{R})}&=&\sqrt{\frac{\mu\pm\sqrt{E^{2}-\Delta_{\text{R}}^{2}}}{t_{\text{SC}}}-k_{y}^{2}}\ .
\end{eqnarray}

The wave vectors for the SCs effectively remain the same as mentioned in AM-SC junction, though to distinguish the left and right leads, we write $u(v)=\sqrt{\frac{1}{2}\left(1\pm\sqrt{1-\frac{\Delta_{\text{L(R)}}^{2}}{E^{2}}}\right)}$. The thermoelectric quasiparticle current is obtained by calculating the scattering coefficients. These can be computed by employing the boundary conditions for the AM-JJ as~\cite{Chen2025, Lu2024}
\begin{widetext}
    \begin{eqnarray}
\psi_{e(h)\text{L}}^{\sigma}|_{x=0^{-}}&=&\psi_{\text{AM}}^{\sigma}|_{x=0^{+}} ,\\
\psi_{\text{AM}}^{\sigma}|_{x=d^{-}}&=&\psi_{e(h)\text{R}}^{\sigma}|_{x=d^{+}} ,\\
\left(t_{0} \tau_{0}-\sigma t_{1} \tau_{z}\right) (-i \partial_{x}\psi_{\text{AM}}^{\sigma})|_{x=0^{+}}-t_{\text{SC}} \tau_{0} (-i\partial_{x} \psi_{e(h)\text{L}}^{\sigma})|_{x=0^{-}}&=&-t_{2} k_{y} \sigma\tau_{z}\psi_{\text{AM}}^{\sigma}|_{x=0} ,\\ 
t_{\text{SC}}\tau_{0}(-i\partial_{x}\psi_{e(h)\text{R}}^{\sigma})|_{x=d^{+}}-\left(t_{0}\tau_{0}-\sigma t_{1} \tau_{z}\right)(-i\partial_{x}\psi_{\text{AM}}^{\sigma})|_{x=d^{-}}&=& t_{2} k_{y} \sigma \tau_{z}\psi_{\text{AM}}^{\sigma}|_{x=d} \ .
 \end{eqnarray}
\end{widetext}
\subsection{Separation of thermoelectric current} \label{appen:evenodd}

The current calculated in the main text for the JJ can contain contributions arising from both SC Cooper pairs and quasiparticles. Hence, in order to identify whether it is the traditional dissipationless Cooper pair current or the dissipative single quasiparticle current, we now separate the current into two parts corresponding to the quasiparticle and Cooper pair contributions. We showcase the total thermoelectric current for our AM-based JJ in Fig.~\ref{L_evenodd_0.8_0.5}(a) for $t_{1}=0.8$ and $t_{2}=0$. The behavior of the thermoelectric current with respect to the SC phase difference $\phi$ satisfies a similar spin-split current behavior as presented in Fig.~\ref{L_vs_phi}(b). This is expected as the total current is the sum of up- and down-spin currents due to the absence of any spin mixing. As the thermoelectric current in JJ may contain both the dissipative (even in $\phi$) and non-dissipative 
(odd in $\phi$) quasiparticle currents, we separate the even and odd components of the total thermoelectric current by reversing the SC phase and show the behavior in Figs.~\ref{L_evenodd_0.8_0.5}(b) and ~\ref{L_evenodd_0.8_0.5}(c), respectively. We find that the total current $\mathcal{L}$ is largely contributed by the even component, i.e., the dissipative current in the junction as presented in Fig.~\ref{L_evenodd_0.8_0.5}(b), whereas the odd current is negligibly small in magnitude compared to the even counterpart. The odd part of the thermoelectric current follows the sinusoidal behavior since it is the nondissipative typical Josephson current [see Fig.~\ref{L_evenodd_0.8_0.5}(c)]. On the other hand, the even thermoelectric current signifies the dissipative current with nonsinusoidal behavior as carried by the quasiparticles in a JJ as shown in Fig.~\ref{L_evenodd_0.8_0.5}(b). Thus, we confirm that the thermoelectric currents presented in the main text for the AM-JJ are indeed dissipative.
\begin{figure}
\includegraphics[width=0.485\linewidth]{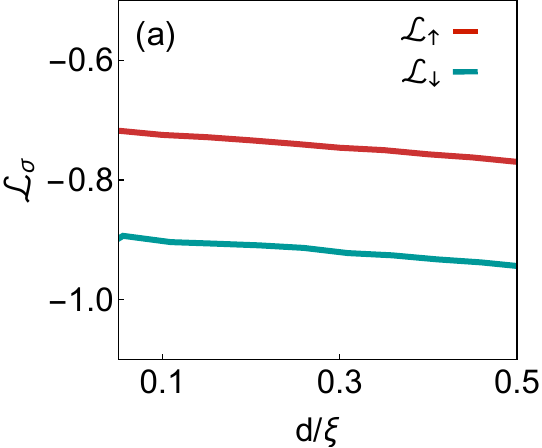}
\includegraphics[width=0.485\linewidth]{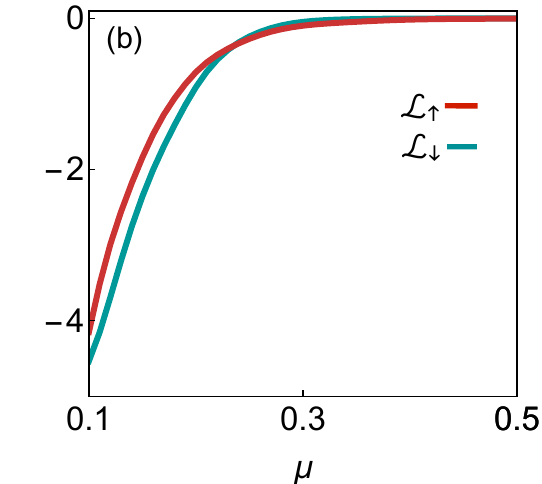}
\caption{(a) The quasiparticle driven thermoelectric current ${L}_{\sigma}$ (in units of $ek_{B}/h$) as a function of the length of the junction $d/\xi$. The other model parameters are considered as $\mu=0.25=250\Delta_{0}$, with $\Delta_{0}=0.001$. (b) The thermoelectric current as a function of the chemical potential $\mu$ of the JJ. We choose the other model parameters as $d=0.1\xi, T=0.3T_{c}$. For both (a) and (b), the AM strengths are considered as $t_{1}=0.8$ and $t_{2}=0$.}
\label{L_vs_d_mu:JJ}
\end{figure}
\subsection{Role of junction length and chemical potential}\label{appen:L_vs_d_mu}
An important aspect of a JJ is the length of the intermediate scatterer region. For the thermoelectric current in a JJ driven by a thermal gradient, it is justified to maintain the length of the JJ, $d$, smaller than the SC coherence length $\xi$. Throughout this work, we have considered a short JJ. 
Here, we present the variation of the thermoelectric current with respect to the length of the intermediate $d$-wave AM. We observe that the thermoelectric current varies very slowly with the junction length associated with soft oscillatory behavior within the short junction regime as depicted in Fig.~\ref{L_vs_d_mu:JJ}(a). We also check the variation of the current with the chemical potential and show it in Fig.~\ref{L_vs_d_mu:JJ}(b). The quasiparticle current is marginally tunable by the chemical potential $\mu$ of the system for both up- and down-spin components, as clearly visible in Fig.~\ref{L_vs_d_mu:JJ}(b). Note that, we have considered $\mu_{\rm AM}=\mu_{\rm SC}$ throughout our article.

\begin{figure}
\includegraphics[width=0.4845\linewidth]{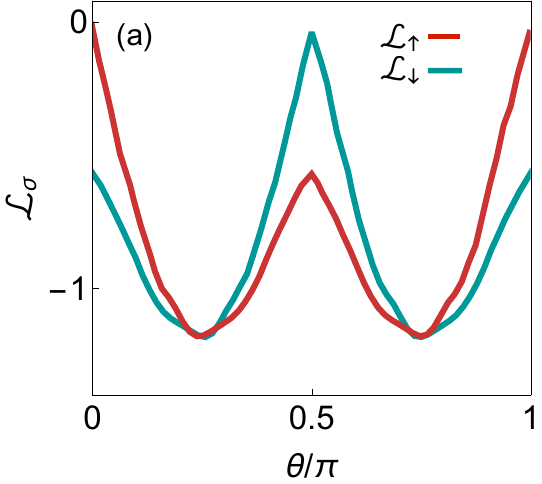}
\includegraphics[width=0.4845\linewidth]{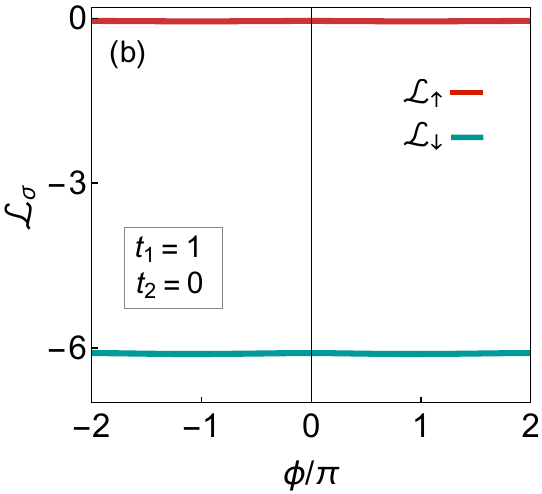}
\caption{(a) Thermoelectric quasiparticle currents $\mathcal{L}_{\sigma}$ (in units of $ek_{B}/h$) with respect to the AM rotational angle $\theta$, considering $\phi=\pi/4$. (b) The spin dependent thermoelectric quasiparticle currents $\mathcal{L}_{\sigma}$ (in units of $ek_{B} /h$) with respect to the superconducting phase difference $\phi$, for $\theta=0$ (i.e. $t_{1}=1$ and $t_{2}=0$) in the JJ. Here, we consider the other parameters of the model Hamiltonian as $\mu=200\Delta_{0}, d=0.1\xi, T=0.4 T_{c}, \Delta_{0}=0.001$.}
\label{L_vs_theta}
\end{figure}

\subsection{Effect of AM rotational angle}\label{appen:L_vs_theta}
The AM term of Eq.~\eqref{Ham:AM_JJ} in the ${k_x, k_y}$ basis can be expressed as
\begin{eqnarray}
   t_{\bold{k},\sigma}^{\text{JJ}}&=& \left(\begin{array}{cc}
k_{x} & k_{y} 
\end{array}\right) 
\left(\begin{array}{cc}
-t_{1}  & t_{2}\\
t_{2} & t_{1}
\end{array}\right) 
\left(\begin{array}{c}
k_{x}  \\ k_{y} 
\end{array}\right)\ .
\end{eqnarray}
Following this matrix representation, we can rewrite the AM Hamiltonian as a function of AM rotational angle $\theta$, by defining the parameters as follows:
\begin{equation}
    t_{1}=\text{cos}(2\theta) \hspace{0.3cm} {\rm{and}} \hspace{0.3cm}
    t_{2}=\text{sin}(2\theta)\ ,
\end{equation}
where $\theta=\frac{1}{2}\text{tan}^{-1}(t_{2}/t_{1})$ denotes the AM angle, which implies the rotation angle of the altermagnetic spin splitting axis, directly linked to the crystal rotational symmetry. In Fig.~\ref{L_vs_theta}, we present the spin-dependent thermoelectric current with respect to the AM rotational angle. The behavior of the current is solely dependent on the crystal rotational symmetry described by $\theta$. Here, $\theta=0, \pi/2, \pi$ imply the $d_{x^2-y^2}$ symmetry of the AM, whereas $\theta=\pi/4,~3\pi/4$ present $d_{xy}$ symmetry. Therefore, the highest current for an individual spin component is found only at specific $\theta$ values following $d_{x^2-y^2}$ symmetry, satisfying $t_{1}=1$ and $t_{2}=0$, as shown in Fig.~\ref{L_vs_t:JJ}(a). Notably, the thermoelectric spin currents presented in Figs.~\ref{L_vs_phi}(a) and ~\ref{L_vs_phi}(b) represent perpendicular Fermi surfaces for the up- and down-spin electrons, but the perpendicular spin splitting axis appears only for $t_{1}=1$ and $t_{2}=0$, i.e., $\theta=0$ (or $\pi/2$ or $\pi$) which we present here in Fig.~\ref{L_vs_theta}(b). Note that, the rest of our paper contains the results following the Hamiltonian of Eq.~\eqref{ham:am} [or Eq.~\eqref{Ham:AM_JJ}] only.

\section{Scattering matrix formalism for AM-based JJ with RSOI} \label{scatmat:JD}
The presence of RSOI in our AM-based JJ allows spin mixing in the system, which is incorporated in the wave function with the additional spin-flip terms as presented in Sec.~\ref{SC_AM_SC:JJ}. In this appendix, we present the wave function for each region
and the boundary conditions required for the RSOI-induced AM-JJ. Apart from spin-conserving normal and Andreev reflection, a spin-up incident particle can undergo a spin-flip ordinary reflection and Andreev reflection due to the Rashba SOC. Hence, we can write the modified wave function for the model heterojunction of Fig.~\ref{fig:AMSC_AM_AMSC}(a) as
\begin{widetext}
\begin{eqnarray}
\psi_{e(h)\text{L}}^{\sigma}&=&\psi_{e\text{L}(h\text{L}),\sigma}^{+}+r_{\text{N}_{e(h)}}^{\sigma} \psi_{e\text{L}(h\text{L}),\sigma}^{-}+r_{\text{A}_{h(e)}}^{\sigma} \psi_{h\text{L}(e\text{L}),\Bar{\sigma}}^{-}+r_{\text{N}_{e(h)}}^{\Bar{\sigma}} \psi_{e\text{L}(h\text{L}),\Bar{\sigma}}^{-}+r_{\text{A}_{e(h)}}^{\Bar{\sigma}} \psi_{h\text{L}(e\text{L}),\sigma}^{-}\ , \\
 \psi_{\text{AM}}^{\sigma}&=&a^{\sigma} \psi_{e(h),\sigma}^{+}+b^{\sigma} \psi_{e(h),\sigma}^{-}+c^{\sigma} \psi_{h(e),\Bar{\sigma}}^{+} + d^{\sigma} \psi_{h(e),\Bar{\sigma}}^{-}+a^{\Bar{\sigma}} \psi_{e(h),\Bar{\sigma}}^{+}+b^{\Bar{\sigma}} \psi_{e(h),\Bar{\sigma}}^{-}+c^{\Bar{\sigma}} \psi_{h(e),\sigma}^{+} + d^{\Bar{\sigma}} \psi_{h(e),\sigma}^{-}\ ,\\
 \psi_{e(h)\text{R}}^{\sigma}&=&t_{ee(hh)}^{\sigma} \psi_{e\text{R}(h\text{R}),\sigma}^{+}+t_{he(eh)}^{\sigma} \psi_{h\text{R}(e\text{R}),\Bar{\sigma}}^{+}+t_{ee(hh)}^{\Bar{\sigma}} \psi_{e\text{R}(h\text{R}),\Bar{\sigma}}^{+}+t_{he(eh)}^{\Bar{\sigma}} \psi_{h\text{R}(e\text{R}),\sigma}^{+}\ ,
\end{eqnarray}
\end{widetext}
where the normal and Andreev reflection amplitudes for the spin-flip counterparts are denoted as $r_{\text{N}_{e(h)}}^{\Bar{\sigma}}$ and $r_{\text{A}_{e(h)}}^{\Bar{\sigma}}$, respectively, and the spin-flip electronlike and holelike  quasiparticle transmission coefficients are given by $t_{ee(he)}^{\Bar{\sigma}}$ and $t_{he(eh)}^{\Bar{\sigma}}$, respectively. The wave functions remain the same as mentioned in Appendix~\ref{scatmat:JJ}.
In the presence of RSOI, the Hamiltonian of Eq.~(\ref{ham:JD}) does not commute with the spin component $\sigma$. Therefore, considering the current conservation along the direction of the thermoelectric current, we calculate the boundary condition for the 
RSOI inserted AM-SC JJ as follows:
\begin{widetext}
    \begin{eqnarray}
\psi_{e(h)\text{L}}^{\sigma}|_{x=0^{-}}&=&\psi_{\text{AM}}^{\sigma}|_{x=0^{+}} \ ,\\
\psi_{\text{AM}}^{\sigma}|_{x=d^{-}}&=&\psi_{e(h)\text{R}}^{\sigma}|_{x=d^{+}}\ ,\\
\left(t_{0} \tau_{0}-\sigma t_{1} \tau_{z}\right) (-i \partial_{x}\psi_{\text{AM}}^{\sigma})|_{x=0^{+}}-t_{\text{SC}} \tau_{0} (-i\partial_{x} \psi_{e(h)\text{L}}^{\sigma})|_{x=0^{-}}+\alpha \sigma_{y} \tau_{z}|_{x=0^{-}}&=&-t_{2} k_{y} \sigma\tau_{z}\psi_{\text{AM}}^{\sigma}|_{x=0} \ ,\\
t_{\text{SC}}\tau_{0}(-i\partial_{x}\psi_{e(h)\text{R}}^{\sigma})|_{x=d^{+}}-\left(t_{0}\tau_{0}-\sigma t_{1} \tau_{z}\right)(-i\partial_{x}\psi_{\text{AM}}^{\sigma})|_{x=d^{-}}+\alpha \sigma_{y} \tau_{z}|_{x=0^{-}}&=& t_{2} k_{y} \sigma \tau_{z}\psi_{\text{AM}}^{\sigma}|_{x=d}\ .
 \end{eqnarray},
\end{widetext}
We use these boundary conditions to solve for the 
scattering amplitudes and calculate the thermoelectric current therein employing Eq.~(\ref{L12}) as discussed in the main text.
\end{appendix}
\bibliography{bibfile}{}
\end{document}